\documentstyle[epsfig]{mn}

\title{Two cosmological models for clusters of galaxies}
\author[Ivan Suhhonenko and Mirt Gramann] 
{Ivan Suhhonenko and Mirt Gramann 
   \\ 
   Tartu Observatory,
       T\~oravere EE-2444, Estonia}
\begin{document}
\maketitle

\let\sec=\section 
\let\ssec=\subsection 
\let\sssec=\subsubsection

\def\kms{\;{\rm km\,s^{-1}}}
\def\kmsmpc{\;{\rm km\,s^{-1}\,Mpc^{-1}}}
\def\hompc{\,h\,{\rm Mpc}^{-1}}
\def\mpcoh{\,h^{-1}\,{\rm Mpc}}
\def\mpc3h{\,h^{3}\,{\rm Mpc^{-3}}}

\begin{abstract}

We investigate the properties of clusters of galaxies in two  
cosmological models using N-body simulations and the Press-Schecter 
(PS) theory. In the first model, 
the initial power spectrum of density fluctuations is in the form 
$P(k) \propto k^{-2}$ at wavelengths $\lambda<120h^{-1}$ Mpc. In the second 
model, the initial linear power spectrum of density fluctuations contains a 
feature (bump) at wavelengths $\lambda \sim 30-60h^{-1}$ Mpc which correspond 
to the scale of superclusters of galaxies. We examine the mass function, 
peculiar velocities, the power spectrum and the correlation function of 
clusters in both models for different values of the density parameter $\Omega_0$ 
and $\sigma_8$ (the rms fluctuation on the $8h^{-1}$Mpc scale). The results 
are compared with observations. We find that in many aspects the power 
spectrum of density fluctuations in the model (2) fits the observed data 
better than the simple power law model (1). In the first model, the mass 
function and peculiar velocities of clusters are consistent with observations 
only if $\Omega_0<0.6$. In the second model, the permitted region in the 
($\Omega_0,\sigma_8$) plane is larger. In this model, the power spectrum of 
clusters is in good agreement with the observed power spectrum of the APM 
clusters. This model predicts that there is a bump in the correlation 
function of clusters at separations $r\sim 20-35h^{-1}$ Mpc. In the future, 
accurate measurements of the cluster correlation function at these distances 
can serve as a discriminating test for this model.

We examine the linear theory predictions for the peculiar
velocities of peaks in the Gaussian field and compare these to the 
peculiar velocities of clusters in N-body simulations. We determine the 
clusters as the maxima of the density field smoothed on the scale
$R\sim 1.5h^{-1}$ Mpc and define their peculiar velocities using the
same smoothing scale as for the density field. The numerical 
results show that in this case the rms peculiar velocities of clusters 
increase with cluster richness. The rms peculiar velocity of small 
clusters is similar to the linear theory expectations, while the rms 
peculiar velocity of rich clusters is higher than that predicted in the 
linear theory ($\sim 18$\% for clusters with a mean intercluster separation 
$d_{cl}=30h^{-1}$ Mpc). 

\end{abstract}


\sec{INTRODUCTION}

The formation of structure in our Universe is one of the most
fascinating problems in cosmology. Usually we believe that galaxies and
clusters of galaxies have developed by gravitational instability out of
small inhomogeneities of the early Universe. The initial field of
density fluctuations $\delta({\bf x},t)$ can be decomposed into its
Fourier components $\delta_{\bf k}(t)$ and expressed in terms of the
power spectrum $P(k)=\langle\vert\delta_{\bf k} \vert^2\rangle$.

\begin{figure*}
\centering
\begin{picture}(300,350)
\includegraphics{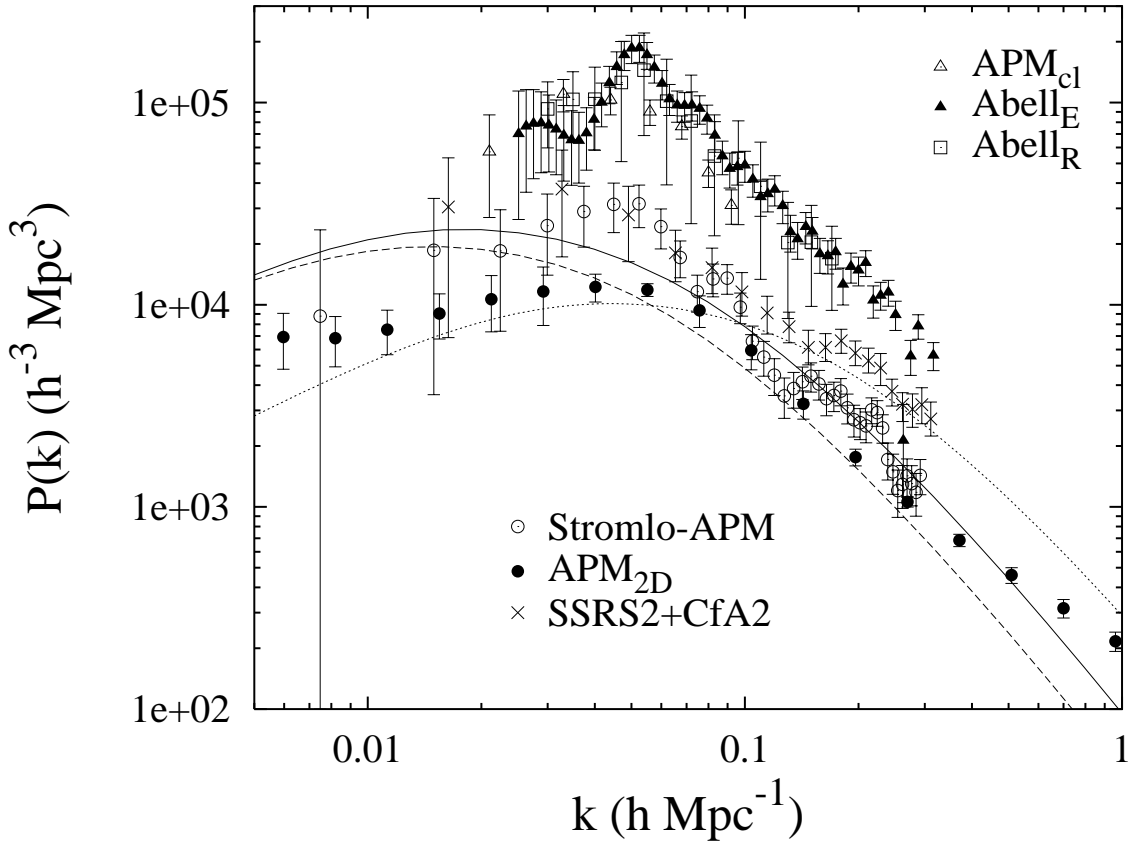}
\end{picture}
\caption {The power spectrum of the distribution of 
galaxies and clusters of galaxies. Filled circles, open circles and 
crosses show the power spectrum of the galaxy distribution in the APM, 
Stromlo-APM and SSRS2+CfA2 surveys, respectively. Filled triangles and 
open squares show the power spectrum of the distribution of Abell clusters
as determined by Einasto et al. (1997a) and Retzlaff at al. (1998),
respectively. Open triangles represent the power spectrum of APM 
clusters. For comparison, we show the linear power spectrum of density
fluctuations in the flat CDM models with $\Omega_0=0.3$ and $h=0.7$ 
(solid line) and $h=0.6$ (dashed line). The dotted line shows the power 
spectrum in the CDM model with $\Omega=1$ and $h=0.5$. The CDM 
models are COBE-normalized.}
\end{figure*}

Figure~1 shows the observed power spectra derived from the distribution of
galaxies in the APM, Stromlo-APM and SSRS2+CfA2 surveys (Baugh \&
Efstathiou 1993; Tadros \& Efstathiou 1996; Costa et al. 1994). 
The power spectrum of the galaxy distribution in the Stromlo-APM
redshift survey peaks at the wavenumber $k=0.052 h$ Mpc$^{-1}$ (or at the
wavelength $\lambda=120 h^{-1}$ Mpc). A similar peak in the
one-dimensional power spectrum of a deep pencil-beam survey was
detected by Broadhurst et al. (1990) and in the two-dimensional power
spectrum of the Las Campanas redshift survey by Landy et al. (1996). 
Available data, however, are insufficient to say whether the peak in 
the Stromlo-APM survey reflects a real feature in the galaxy distribution. 
It is likely that the decline in the power spectrum at wavenumbers 
$k\le0.052h$ Mpc$^{-1}$ is partly due to the effects of the uncertainty 
in the mean number density of optical galaxies (see Tadros \& Efstathiou 
1996 for a discussion of this effect). However, independent evidence for 
the presence of a preferred scale in the Universe at about $120h^{-1}$ Mpc 
comes from an analysis of the distribution of galaxy clusters. 
Figure~1 shows the power spectrum of the distribution of the Abell-ACO 
clusters as determined by Einasto et al. (1997a) and Retzlaff et al. (1998), 
and the power 
spectrum of the APM clusters as measured by Tadros, Efstathiou \& Dalton 
(1998). The power spectrum of the distribution of the Abell-ACO clusters has 
a well-defined peak at the same wavenumber, $k_0=0.052h$ Mpc$^{-1}$, as 
the power spectrum of galaxies in the Stromlo-APM survey. For wavenumbers
$k>k_0$, the shape of the clusters' power spectrum is similar to the
shape of the power spectrum for galaxies in the Stromlo-APM survey.
This comparison suggests that the peak observed in the power spectrum
of the Stromlo-APM redshift survey is a real feature in the 
distribution of galaxies (see Gramann (1998) for a more detailed 
discussion of the observed power spectra in different galaxy surveys). 

Cosmological models based on collisionless dark matter (e.q. cold dark
matter (CDM)) and adiabatic fluctuations, when combined with power-law 
initial power spectra, predict smooth power spectra of density 
fluctuations at $z\sim 10^3$. Figure~1 shows the power spectra of density 
fluctuations predicted in the flat CDM models with the density parameter 
$\Omega_0=0.3$ and the normalized Hubble constant $h=0.6$ and $h=0.7$. For 
comparison, we show in Figure~1 the power spectrum predicted in the CDM 
model with $\Omega=1$ and $h=0.5$. We have used the transfer function 
derived by Bardeen et al. (1986) and Sugiyama (1995), and the COBE 
normalization derived by Bunn and White (1997). The observed power spectra 
of galaxies and clusters of galaxies are not consistent with CDM-type models 
(see also e.g. Peacock 1997; Einasto et al. 1997a; Tadros, Efstathiou \& 
Dalton 1998). The baryonic acoustic oscillations in adiabatic models may 
explain the observed excess only if currently favored determinations 
of cosmological parameters are in substantial error (e.g. the density 
parameter $\Omega_0 < 0.2h$) (Eisenstein et al. 1998). One possible 
explanation for the presence of a peak in the power spectrum is an 
inflationary scenario with a scalar field whose potential has a localized 
feature around some value of the field (Starobinsky 1992; Lesgourgues, 
Polarski \& Starobinsky 1998).

\begin{figure*}
\centering
\begin{picture}(300,350)
\includegraphics{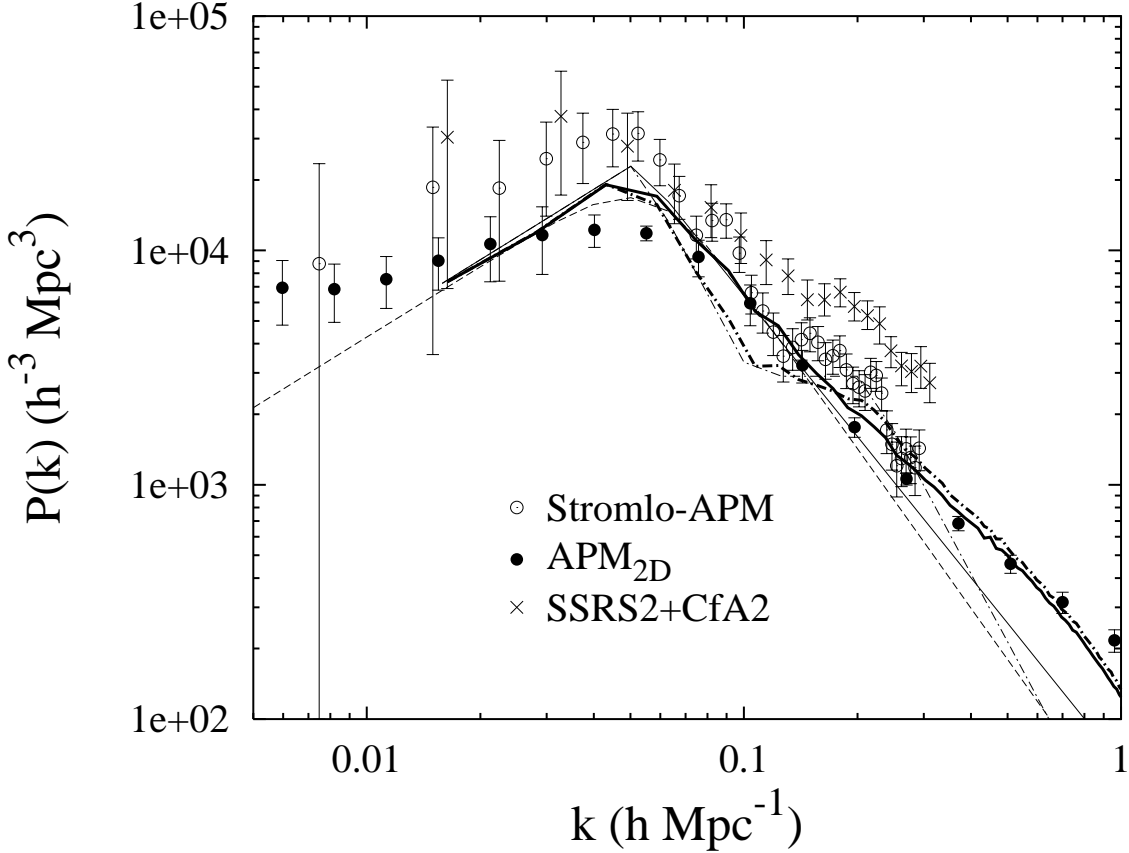}
\end{picture}
\caption {The linear and non-linear power spectra of density
fluctuations in the model (1) (solid lines) and in the model (2)
(dot-dashed lines). The heavy curves show the results of
the N-body simulations and the light curves the corresponding linear
power spectra. The dashed line shows the linear power spectrum
derived by Peacock (1997). Filled circles, open circles 
and crosses show the power spectrum of the galaxy distribution in 
the APM, Stromlo-APM and SSRS2+CfA2 surveys, respectively.}
\end{figure*}

In this paper we study the properties and spatial distribution of
galaxy clusters in two cosmological models which start from the
observed power spectra of the distribution of galaxies. In the first
model we assume that the initial linear power spectrum of density
fluctuations in the universe at $z\sim 10^3$ is of the form 
$$
P(k)=\cases {P(k_0)(k/k_0), &if $k<k_0$;\cr   
P(k_0) (k/k_0)^{-2}, &if $k > k_0$ , \cr}
\eqno(1)
$$
where $k_0=0.052h$ Mpc$^{-1}$ and $P(k_0)=3.71 \times 10^4 \sigma_8^2
h^{-3}$ Mpc$^3$. The $\sigma_8$ is the rms mass fluctuation on the
$8h^{-1}$ Mpc scale. Gramann (1998) used the function (1) to
recover the power spectrum in the Stromlo-APM redshift survey. In the 
second model we assume that the linear power spectrum contains a 
primordial feature at wavenumbers $k\sim 0.1 - 0.2 h$ Mpc$^{-1}$ 
($\lambda \sim 30-60h^{-1}$ Mpc) which correspond to the scale of
superclusters, and
$$
P(k)=\cases {P(k_0)(k/k_0), &if $k<k_0$;\cr   
P(k_0) (k/k_0)^{-3}, &if $k_0<k < k_1$ ; \cr
P(k_1), & if $k_1<k<k_2$; \cr
P(k_1) (k/k_0)^{-3}, & if $k>k_2$, }
\eqno(2)
$$
where $k_0=0.052h$ Mpc$^{-1}$, $k_1=0.1h$ Mpc$^{-1}$, $k_2=0.2h$ Mpc$^{-1}$,
$P(k_0)=3.34 \times 10^4 \sigma_8^2 h^{-3}$ Mpc$^3$ and $P(k_1)=4.07
\times 10^3 \sigma_8^2 h^{-3}$ Mpc$^3$. We assume also that the initial
density fluctuation field in the Universe is a Gaussian field. In this
case the power spectrum provides a complete statistical description of 
the field.
 
Figure~2 demonstrates the linear and nonlinear power spectra of density 
fluctuations in the model (1) for $\sigma_8=0.8$ and in the model (2) 
for $\sigma_8=0.84$. To examine the nonlinear evolution of density
fluctuations we have used N-body simulations. The parameters of the
simulations are given in Section 3. We see that at wavenumbers 
$k>0.2h^{-1}$ Mpc, the nonlinear power spectra in models (1) and (2) are 
very similar. Figure~2 shows also the power spectra of the galaxy clustering 
in the APM, Stromlo-APM and SSRS2+CfA2 surveys (Baugh \& Efstathiou 1993; 
Tadros \& Efstathiou 1996; Costa et al. 1994). The spatial distribution
of galaxies in the models (1) and (2) depends on the relation between 
galaxies and matter density, and further study is needed to study galaxies 
in these models. Figure~2 demonstrates that if we assume a linear
bias between galaxies and density, the inital power spectra given by
equations (1) and (2) are in good agreement with the observed power 
spectra of galaxies. For comparison, we show in Figure~2 the linear power 
spectrum derived from the observed galaxy power spectra by Peacock (1997) 
(we have used his eq. [34] with parameters given in eq. [35]). The initial 
power spectrum derived by Peacock (1997) is similar to the function (1). 

In this paper we examine the evolution of the mass function, peculiar 
velocities, the power spectrum and the correlation function of galaxy 
clusters in models (1) and (2) for different values of the density parameter 
$\Omega_0$ and $\sigma_8$. To study the mass function of clusters of galaxies,
we use the Press-Schechter (1974) formalism. To investigate peculiar 
velocities and the spatial distribution of galaxy clusters, we use N-body 
simulations. 

In the simulations we use the equations of motion for the model with 
$\Omega=1$. According to the equations of motion, expressed in terms of 
the linear growing mode, the evolution of a pressureless fluid in an 
expanding universe is almost independent of the density parameter 
$\Omega_0$ and of the cosmological constant $\Lambda$ (e.g. Gramann 1993; 
Nusser \& Colberg 1998). Nusser \& Colberg (1998) used high resolution 
N-body simulations to investigate the effect of changing the cosmological 
background on the evolution of fluctuations and demonstrated that once 
the initial density fluctuation field is evolved to a given amplitude 
(e.g. $\sigma_8 \sim 0.7$) and smoothed on scales $R\ge 1h^{-1}$ Mpc, 
it is almost insensitive to the cosmological background. The smoothed 
nonlinear velocity field scales with the linear velocity growth factor, 
$f(\Omega_0) \approx \Omega_0^{0.6}$, just as it does in the linear theory. 
Therefore, if clusters represent the maxima of the density field which is 
smoothed on scales $R \sim 1-2h^{-1}$ Mpc, their spatial distribution in 
real space is not sensitive to $\Omega_0$ and $\Lambda$. 

Observations provide the distribution of clusters in the redshift space, which 
is distorted due to peculiar velocities of clusters. In order to study 
peculiar velocities of galaxy clusters and their distribution in the redshift 
space in the models with different $\Omega_0$, we determine the velocities 
of clusters in the simulations with $\Omega=1$ and assume that peculiar 
velocities of galaxy clusters, as the whole velocity field, are proportional 
to the growth factor $f(\Omega_0)$. The linear velocity growth factor depends 
very weakly on the cosmological constant (e.g. Lahav et al. 1991) and in 
this paper we neglect this dependence. 

This paper is organized as follows. In Section 2 we study the mass
function of clusters of galaxies in our models and compare the
results with observations. In Section 3 we examine peculiar velocities 
of galaxy clusters. In Section 4 we investigate the redshift-space power 
spectrum of clusters and in Section 5 we study the correlation function of 
clusters. Section 6 summarizes the main results. 

A Hubble constant of $H_0=100h$ km s$^{-1}$Mpc$^{-1}$ is used 
throughout this paper. 

\sec{THE MASS FUNCTION OF CLUSTERS OF GALAXIES}

To investigate the mass function of clusters we use the Press-Schechter 
(1974, PS) approximation. The PS mass function has been compared with
N-body simulations (Efstathiou et al. 1988; White, Efstathiou \& Frenk 
1993; Lacey \& Cole 1994; Eke, Cole \& Frenk 1996; Borgani et al. 1997a) 
and has been shown to provide an accurate description of the abundance 
of virialized halos of cluster size. In the PS approximation the number 
density of clusters with the mass between $M$ and $M+dM$ is given by
$$
n(M) dM = - \sqrt{{2 \over \pi}} {\rho_b \over M} 
{\delta_t \over \sigma^2(M)} {d\sigma(M)\over dM}  
\exp \left[-{\delta_t^2 \over 2\sigma^2(M)}\right]  dM .  
\eqno(3)
$$
Here $\rho_b$ is the mean background density and $\delta_t$ is the linear
theory overdensity for a uniform spherical fluctuation which is now
collapsing; $\delta_t=1.686$ for $\Omega=1$, with a weak
dependence on $\Omega_0$ for flat and open models (e.g. Eke et al.
1996). The function $\sigma(M)$ is the rms linear density fluctuation at the
mass scale $M$. We will use the top-hat window function.
For the top-hat window, the mass $M$ is related to the window radius
$R$ as $M=4\pi\rho_b R^3/3$. In this case, the number density of clusters 
of mass larger than $M$ can be expressed as
$$
n_{cl}(>M) = \, {\int_M^{\infty}} n(M') dM' = 
$$
$$ 
= - {3 \over (2\pi)^{3/2}} {\int_R^{\infty}} { \delta_t \over \sigma^2(r)} 
{d\sigma(r)\over dr} \exp\left[-{\delta_t^2 \over 2\sigma^2(r)}\right] 
{dr \over r^3}  .   
\eqno(4)
$$

\begin{figure*}
\centering
\begin{picture}(300,490)
\includegraphics{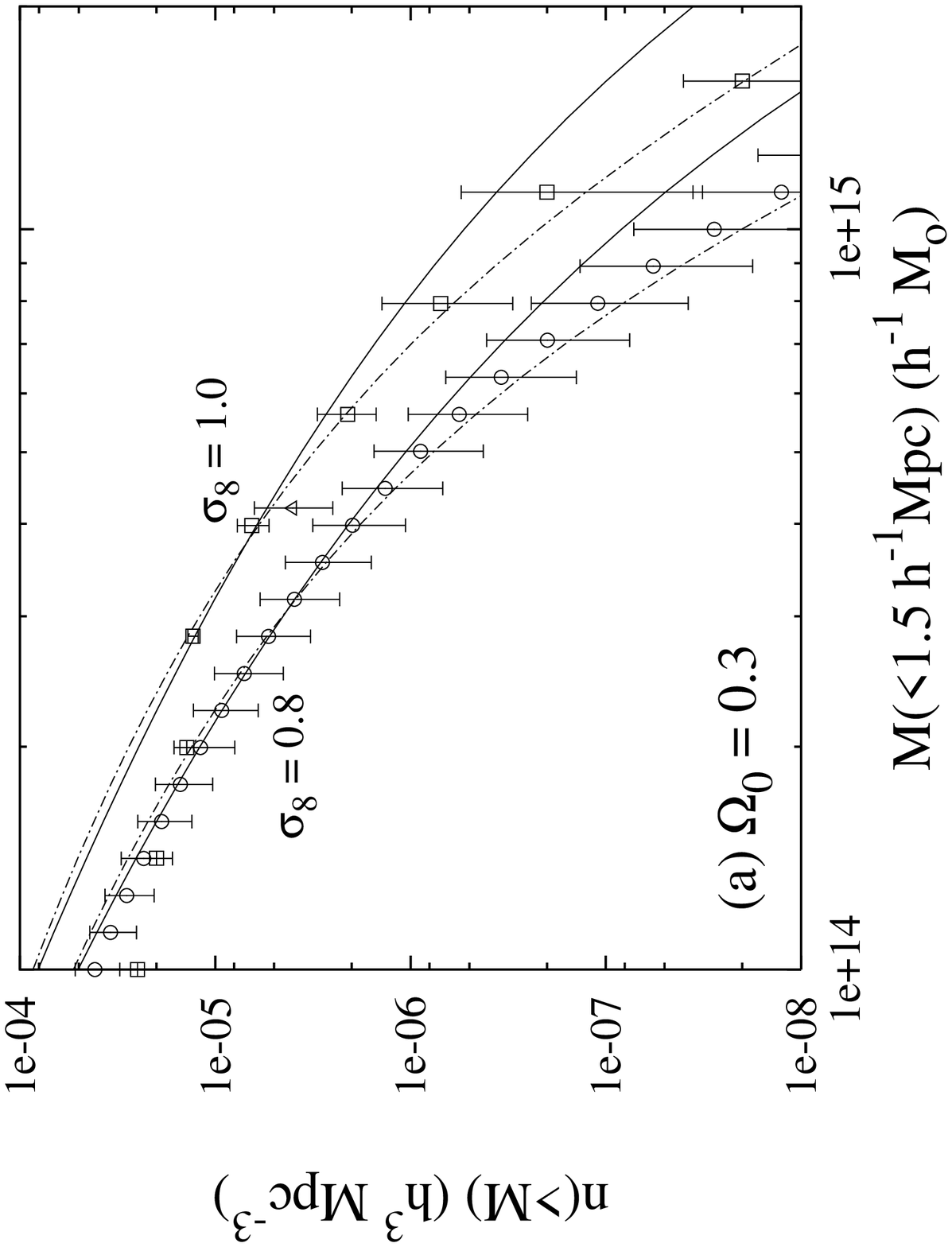}
\includegraphics{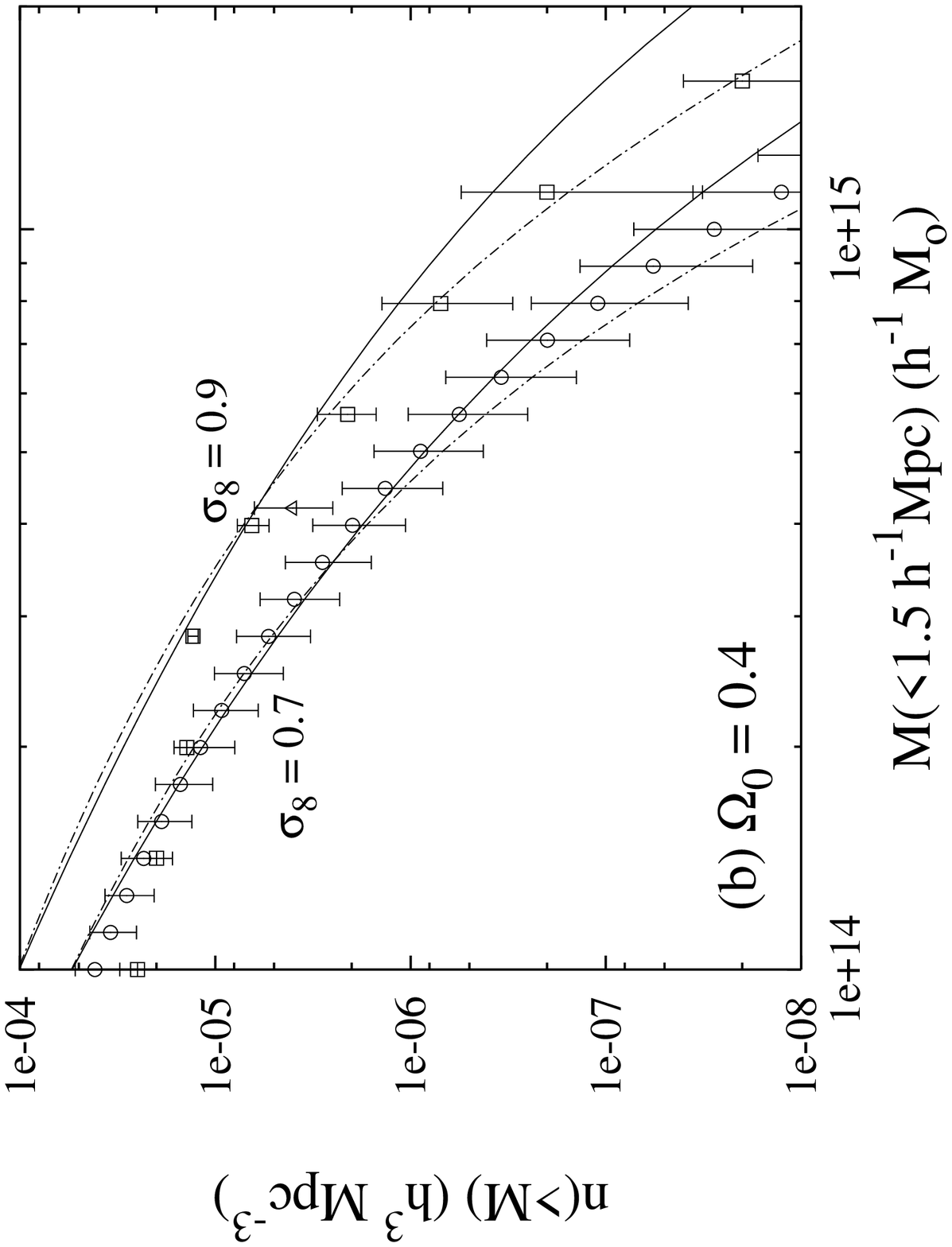}
\end{picture}
\caption {The cluster mass function in the model (1) (solid
lines) and in the model (2) (dot-dashed lines). (a) The density parameter 
$\Omega_0=0.3$. The mass function is shown for $\sigma_8=0.8$ and for
$\sigma_8=1.0$. (b) The density parameter $\Omega_0=0.4$. The mass function 
is shown for $\sigma_8=0.7$ and for $\sigma_8=0.9$. Open circles and 
squares show the mass function of galaxy clusters derived by Bahcall 
and Cen (1993) and by Girardi et al. (1998), respectively. The open 
triangle describes the result obtained by White, Efstathiou \& 
Frenk (1993).}
\end{figure*} 

Figure~3 shows the cluster mass function for the power spectra (1) and (2). 
We investigated the cluster masses within a $1.5h^{-1}$ Mpc radius
sphere around the cluster center. This mass $M_{1.5}$, is related to the 
window radius $R$ as 
$$
R=8.43 \Omega_0^{0.2\alpha \over 3-\alpha} \left[{M_{1.5} \over 6.99
\times 10^{14} \Omega_0 h^{-1} M_{\odot}}\right]
^{1 \over 3-\alpha} (h^{-1} {\rm Mpc})  .
\eqno(5)
$$
Here the parameter $\alpha$ describes the cluster mass profile, 
$M(r) \sim r^\alpha$, at radii $r\sim 1.5h^{-1}$ Mpc. Numerical 
simulations and observations of clusters indicate that the parameter 
$\alpha \approx 0.6-0.7$ for most of clusters  (Navarro, Frenk \& 
White 1995; Carlberg, Yee \& Ellingson 1997). In this paper we use a 
value $\alpha=0.65$. 

The number density of massive clusters is a sensitive function of
$\Omega_0$ and $\sigma_8$. Figure~3a shows the mass function 
for the open models with $\Omega_0=0.3$, $\sigma_8=0.8$ and
$\sigma_8=1.0$. In Figure~3b we present the mass function for the
open models with $\Omega_0=0.4$, $\sigma_8=0.7$ and $\sigma_8=0.9$. 
The cluster abundances in models (1) and (2), when compared at the
same values of $\Omega_0$ and $\sigma_8$, are very similar for smaller 
cluster masses, $M_{1.5} \leq 4 \times 10^{14} h^{-1} M_{\odot}$. For 
larger masses 
the mass function in the model (2) is steeper than in the model (1).
We investigated also the cluster abundances for the flat models with 
slightly larger values of $\delta_t$ derived by Eke et al. (1996), and
found that for a given $\Omega_0$ and $M_{1.5}$, the cluster abundance 
for the flat model is $\sim 10\%$ smaller than for the open model.    

Figure~3 shows also the mass function of clusters of galaxies derived
by Bahcall and Cen (1993, BC) and by Girardi et al. (1998, G98). BC 
used both optical and X-ray observed properties of clusters to determine 
the mass function of clusters. The function was extended towards the faint 
end using small groups of galaxies. G98 determined the mass function 
of clusters by using virial mass estimates for 152 nearby 
Abell-ACO clusters including the new ENACS data (Katgert et al. 1998). The 
mass function derived by G98 is somewhat larger than the mass function 
derived by BC, the difference being larger at larger masses (see Figure~3). 
We find that the models with $\Omega_0=0.3$, $\sigma_8=0.8$ and 
$\Omega_0=0.4$, $\sigma_8=0.7$ provide good match to the mass function 
derived by BC, whereas models with $\Omega_0=0.3$, $\sigma_8=1.0$ and 
$\Omega_0=0.4$, $\sigma_8=0.9$ are in good agreement with the data derived 
by G98.   

Let us consider the amplitude of the mass function of galaxy clusters at 
$M_{1.5} = 4\times 10^{14} h^{-1} M_{\odot}$. For this mass, the cluster 
abundances derived by BC and G98 are $n(>M)=(2.0 \pm 1.1)\times 10^{-6} 
h^3$ Mpc$^{-3}$ and $n(>M)=(6.3\pm 1.2) 10^{-6} h^3$ Mpc$^{-3}$, 
respectively. By analysing X-ray properties of clusters, White, Efstathiou 
\& Frenk (1993) found that the number density of clusters with mass 
$M_{1.5} \approx 4.2 \times 10^{14} h^{-1} M_{\odot}$ is 
$n(>M)=4 \times 10^{-6} h^3$ Mpc$^{-3}$. Figure~4 shows the limits for  
$\sigma_8 \Omega_0^{0.6}$, assuming that the mass function 
of galaxy clusters at $M_{1.5} = 4\times 10^{14} h^{-1} M_{\odot}$ is in 
the range $(2-6.5) \times 10^{-6} h^3$ Mpc$^{-3}$. These limits are similar 
in both our models. For $\Omega_0=0.3$ and $\Omega_0=0.4$, we find that $
\sigma_8=0.90\pm 0.12$ and $\sigma_8=0.80 \pm 0.09$, respectively. For 
$\Omega=1$, $\sigma_8=0.56 \pm 0.05$. These limits for $\sigma_8$ are very 
similar to the limits derived by Eke et al. (1996) for the CDM models
by analyzing X-ray temperatures of clusters.

\begin{figure*}
\centering
\begin{picture}(300,490)
\includegraphics{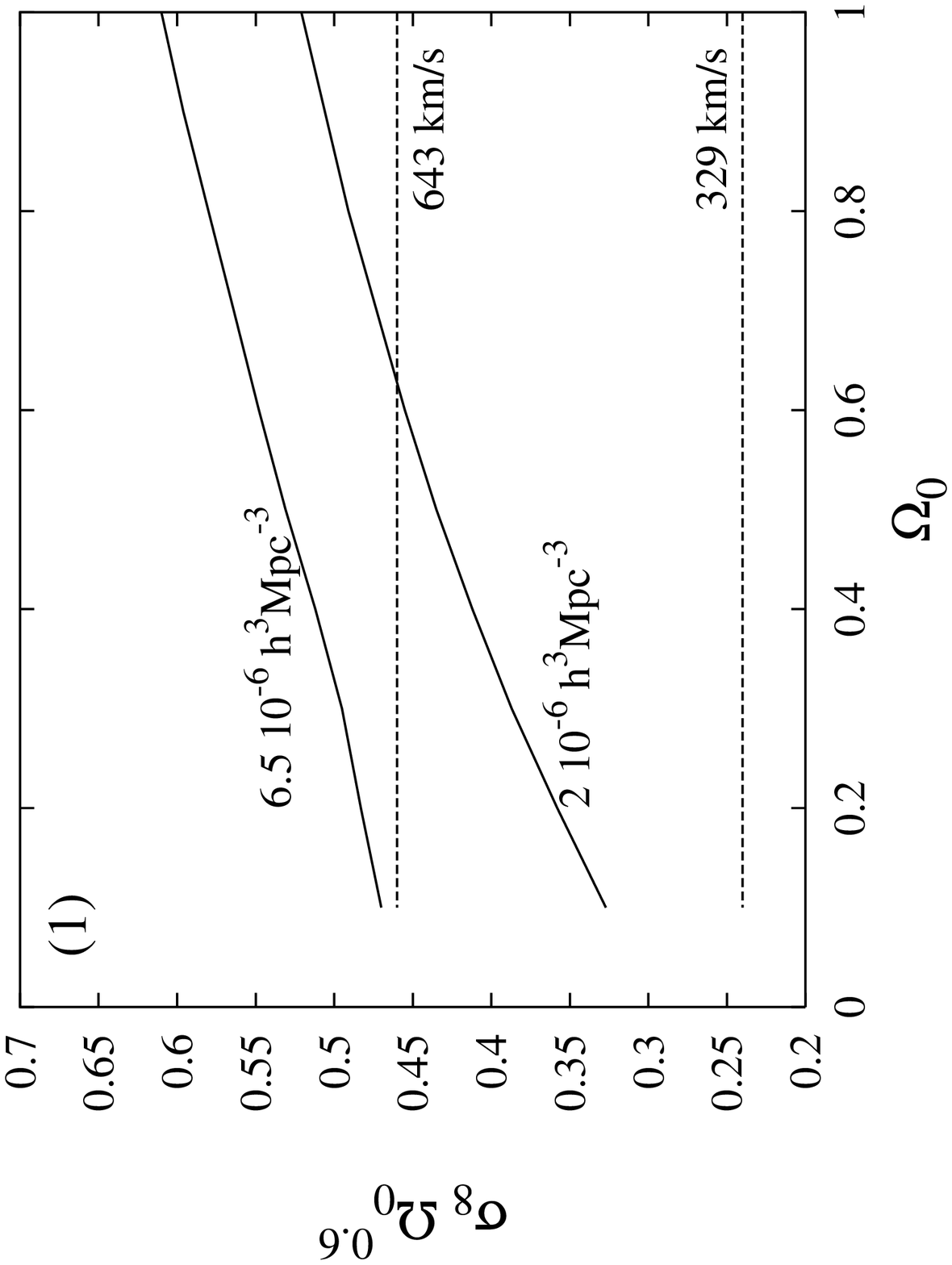}
\includegraphics{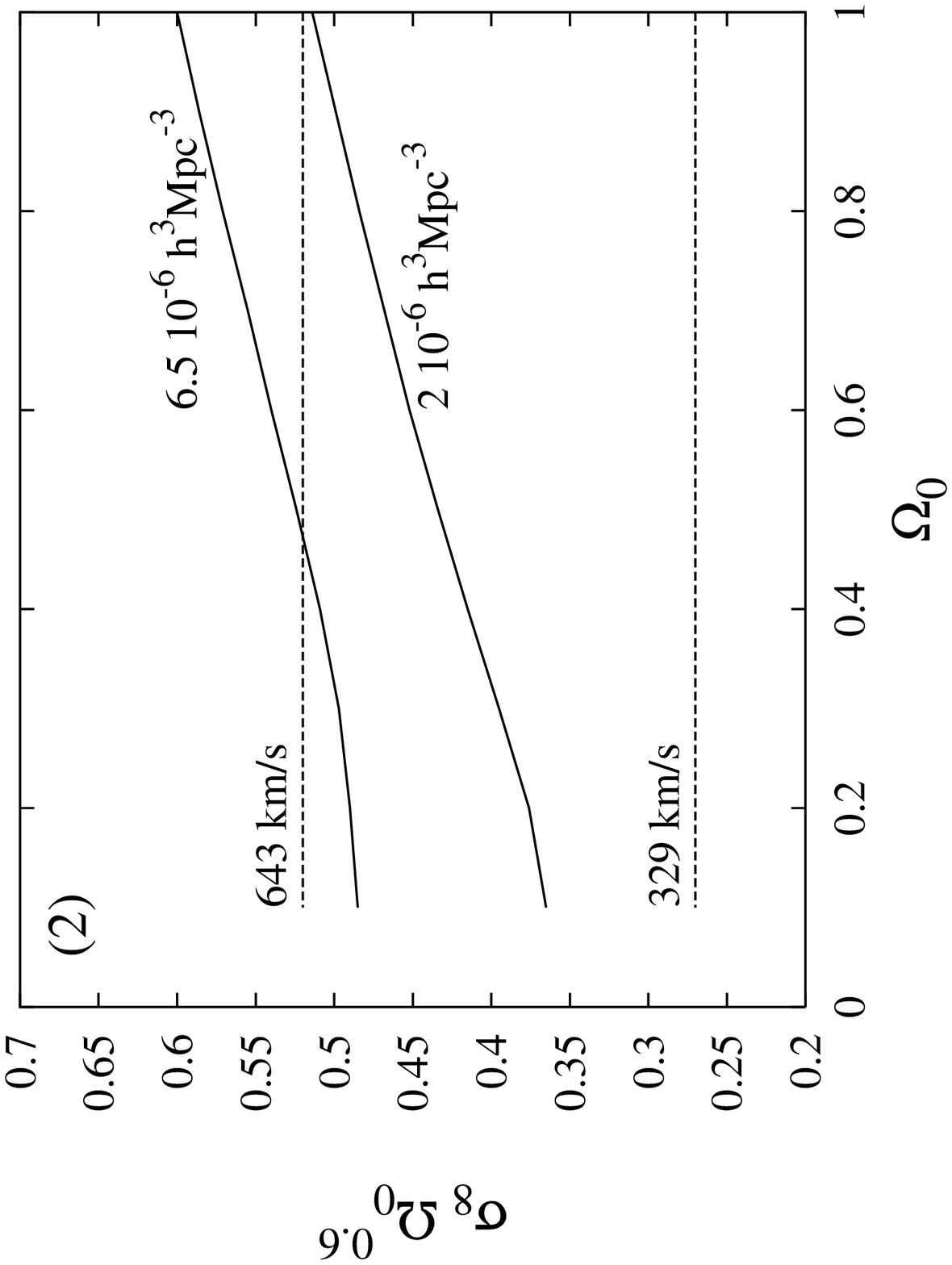}
\end{picture}
\caption {The limits for $\sigma_8 \Omega_0^{0.6}$ in the model (1) (upper
panel) and in the model (2) (lower panel). Solid lines show the
constraints obtained by studying the mass function of clusters and
dashed lines show the constraints obtained by analyzing the peculiar
velocities of clusters.}
\end{figure*} 

\sec{PECULIAR VELOCITIES OF CLUSTERS OF GALAXIES}

To investigate peculiar velocities of galaxy clusters and their
spatial distribution we used N-body simulations. The simulations examined 
in this paper were created using the particle-mesh code described by 
Gramann (1988). We investigated the evolution of ${256}^3$ particles on 
a ${256}^3$ grid, the comoving box size was $L = 384 h^{-1}$ Mpc. 
Clusters were determined in the simulations at the moment when 
$\sigma_8=0.8$ and $\sigma_8=0.84$ in the models (1) and (2), respectively.  

Clusters were selected in the simulations as maxima of the density field 
that was determined on a $256^3$ grid using the CIC-scheme. To determine 
peculiar velocities of clusters, we determined the peculiar velocity 
field on a $256^3$ grid using the CIC-scheme and found the 
peculiar velocities at the grid points were the clusters had been identified. 
The clusters were then ranked according to their density and we selected 
$N_{cl}=(L/d_{cl})^3$ highest ranked clusters to produce cluster catalogs 
with a mean intercluster separation $d_{cl}=10 - 100 h^{-1}$ Mpc. 
For comparison, the number density of the observed APM clusters 
and Abell clusters is $n_{cl}\sim 3.4 \times 10^{-5}h^{3}$ Mpc$^{-3}$ 
($d_{cl}\sim 31 h^{-1}$Mpc) and $n_{cl}\sim 2.5 \times 10^{-5}h^{3}$ 
Mpc$^{-3}$ ($d_{cl}\sim 34h^{-1}$ Mpc), respectively (Dalton et al. 1994, 
Einasto et al. 1997b, Retzlaff et al. 1998).

It is difficult to follow the evolution of rich clusters by using N-body 
simulations, as it requires simulations with a very large dynamical
range to identify correctly a sufficient number of clusters. 
The grid size in our N-body simulations is $R_g=1.5h^{-1}$ Mpc and
the cluster centers in the simulations are at least $3h^{-1}$ Mpc apart. 
This could cause clusters to merge prematurely,
and to have some effect on their properties. However, in order to increase
the resolution of the simulations we must increase the number of
test particles and grid points, or have to follow the evolution of 
clusters in a smaller box. While the first possibility is technically difficult, 
in the latter case the number of rich clusters becomes too small to get
statistically reliable results. Taking into account the requirements on
the number of clusters and on the resolution together with the fact of fixed 
computer resources we decided to use a box size $L=384h^{-1}$ Mpc and  
a grid size $R_g=1.5h^{-1}$ Mpc. 

To determine the rms peculiar velocities of clusters, we
used the equation 
$$
v_{cl}^2=v_s^2+v_L^2=
{1 \over N_{cl}} \sum_{i=1}^{N_{cl}} \, v_i^2 \,\, + v_L^2 \,,
\eqno(6)
$$
where the parameter $v_s$ describes the dispersion of the cluster 
velocities, $v_i$, derived from the simulation and the parameter $v_L$
is given by 
$$
v_L^2 = f^2(\Omega_0) H_0^2 
\left[{1 \over 2\pi^2} \int_0^{\infty} P(k) dk -
{1 \over L^3} \sum_{{\bf k_s}} \, {P_s({\bf k}) \over k^2}\right]. 
\eqno(7)
$$
The last term in this expression is a discrete sum over the linear modes
in the simulation, $\bf k_s$, with the power per mode, $P_s({\bf k})$, as 
actually used in the simulation. We found that for clusters with a mean 
separation $d_{cl}\geq 30h^{-1}$ Mpc, $v_L^2/v_s^2 \sim 4.9$\% and 
$v_L^2/v_s^2 \sim 3.7$\%, in models (1) and (2), respectively.

\begin{figure*}
\centering
\begin{picture}(300,480)
\includegraphics{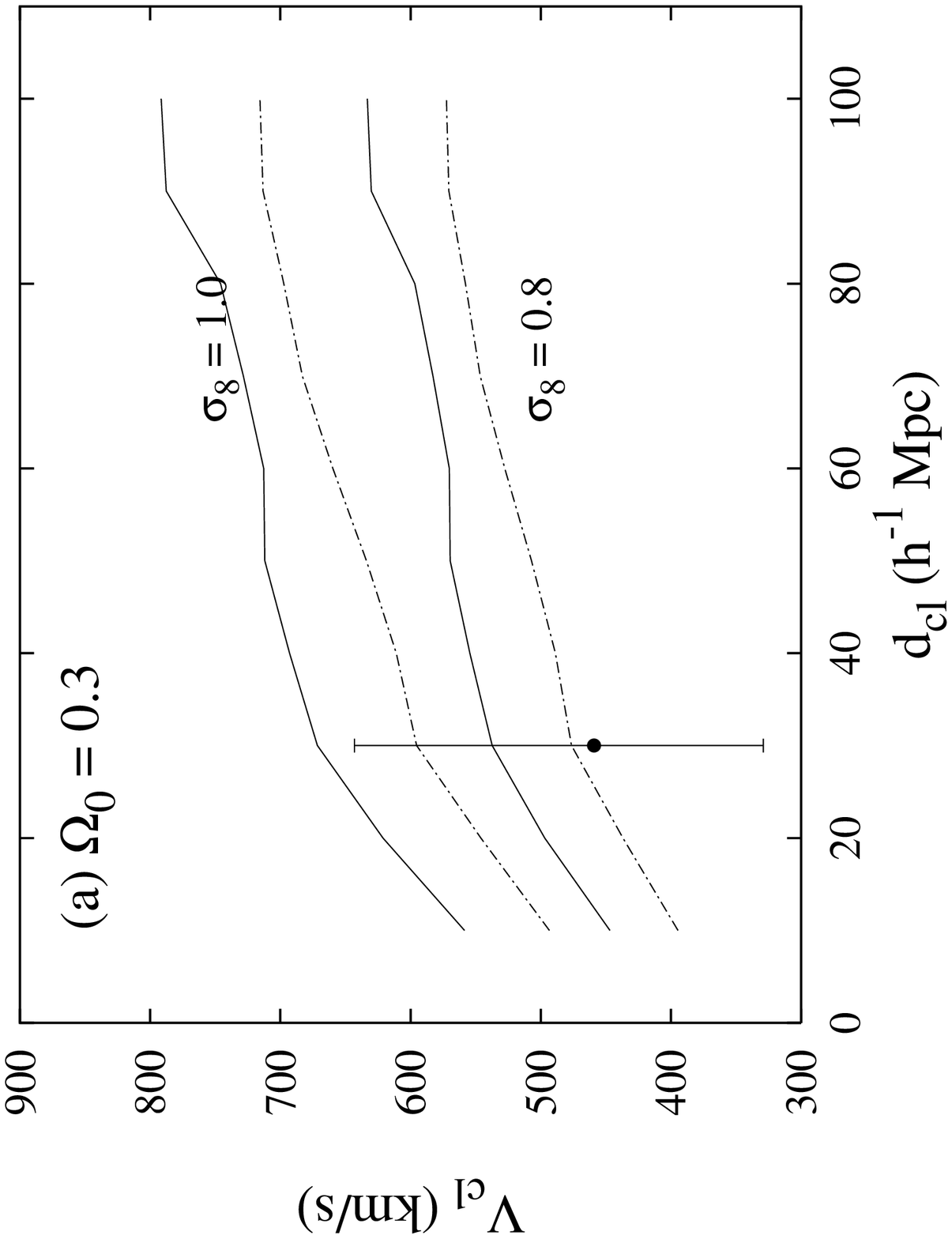}
\includegraphics{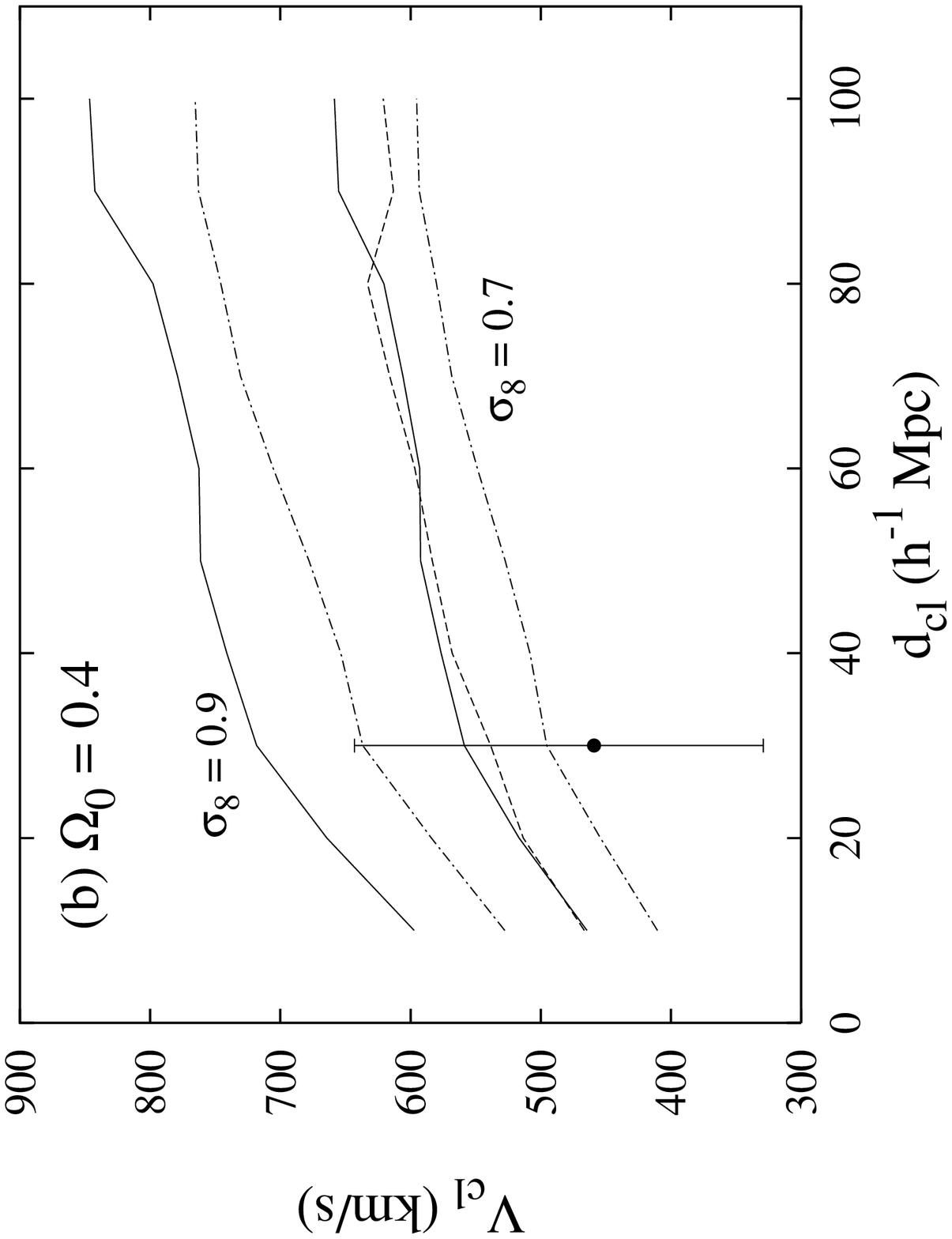}
\end{picture}
\caption {The rms peculiar velocities of clusters in the model (1) 
(solid lines) and in the model (2) (dot-dashed lines). The 
rms velocities are shown for different values of the mean cluster separation,
$d_{cl}$. (a) The density parameter $\Omega_0=0.3$. The rms velocities 
are shown for $\sigma_8=0.8$ and for $\sigma_8=1.0$. 
(b) The density parameter $\Omega_0=0.4$. The rms velocities are shown for 
$\sigma_8=0.7$ and for $\sigma_8=0.9$. For $\sigma_8=0.7$, we show the 
rms cluster velocities determined by simulations when $\sigma_8=0.5$ 
(dashed line) and $\sigma_8=0.8$ (solid line). 
The filled circle shows the observed rms peculiar velocity of galaxy clusters.}
\end{figure*}

Figure~5 shows the rms peculiar velocities of clusters, $v_{cl}$, in  
models (1) and (2), for the same values of $\Omega_0$ and 
$\sigma_8$ as in Figure~3. The rms peculiar velocities of clusters 
in the model (2) are $\sim 12$\% smaller than in the model (1), when 
compared at the same values of $\Omega_0$ and $\sigma_8$. Figure~5 shows 
the rms peculiar velocities for different values of the mean 
cluster separation, $d_{cl}$. We see that the rms peculiar velocities of 
clusters increase with cluster richness. In both models studied, the rms 
peculiar velocity of very massive clusters with an intercluster separation 
$d_{cl}=80h^{-1}$ Mpc is $\sim 25$\% higher than the rms velocity of the 
clusters with a separation $d_{cl}=20h^{-1}$ Mpc. 

We note that this effect is sensitive to the agorithms used to 
define clusters and to determine their peculiar velocities. 
When we determined the clusters using the standard friends-of-friends 
algorithm (FOF) and defined the peculiar velocity of each cluster to be 
the mean peculiar velocity of all the particles within the cluster, we found
that the rms peculiar velocities of clusters decrease with cluster richness.
This is because with the FOF algorithm we identify almost the same objects 
as with the algorithm used throughout this paper, but the sizes of objects 
are different --- poor clusters are smaller and rich clusters are larger. 
The peculiar velocity of the cluster, as the mass of the cluster, depends on 
the size of the cluster. The larger the smoothing region, the smaller the 
velocity. 

In this paper we have determined the clusters as the maxima of the density
field smoothed on the scale $R \sim 1.5h^{-1}$ Mpc and have defined their
peculiar velocities using the same smoothing scale as for the density 
field. In this way, the sizes of all clusters are the same and do not depend 
on the richness of the cluster. Our study shows that in this case the
rms peculiar velocities of clusters increase with cluster richness.
In other words, the rms peculiar velocities of peaks increase with 
the height of the peaks. We found that this result is not very sensitive 
to the power spectrum of density fluctuations. We tested also two other 
models with different power spectra ($n=-1$ and a sCDM model) and 
found that the rms peculiar velocity increases with cluster richness 
similary in all models studied.

Our result is consistent with the result found by Colberg et al. (1998) 
for superclusters. Richer clusters are clustered more strongly and,
therefore, in superclusters the rms mass of clusters is larger than for 
isolated clusters. Colberg et al. (1998) studied the peculiar velocities of 
clusters and found that the rms velocities for clusters which are members of 
superclusters are about 20\% to 30\% larger than those for isolated clusters. 
Therefore, more massive clusters in superclusters move faster than 
less massive clusters outside superclusters. 

To examine the rms peculiar velocities of clusters for different moments 
between $\sigma_8=0.7$ to $\sigma_8=1.0$, we started from the cluster 
velocities at the moment $\sigma_8=0.8$ and $\sigma_8=0.84$, in models (1) 
and (2), respectively, and used the linear scaling, 
$v_{cl} \sim \sigma_8$. We tested this scaling in the model (1), by comparing 
the cluster velocities in N-body simulations at the moments when 
$\sigma_8=0.5$ and $\sigma_8=0.8$. The results are presented in Figure~5b. 
For $\Omega_0=0.4$ and $\sigma_8=0.7$, we show the rms cluster peculiar 
velocities determined by using N-body simulations of clusters when 
$\sigma_8=0.5$ (dashed line) and $\sigma_8=0.8$ (solid line). We see that the differences 
for clusters with $d_{cl}<80h^{-1}$ Mpc are very small. The difference is 
somewhat larger for massive clusters with a mean separation 
$d_{cl}>80h^{-1}$ Mpc ($\sim 5$\%). Therefore, during the evolution 
between $\sigma_8=0.5$ and $\sigma_8=0.8$, cluster velocities 
evolve almost as expected by the linear approximation. 

The observed rms peculiar velocity of galaxy clusters was investigated 
in several recent papers (e.g. Bahcall, Gramann \& Cen 1994, 
Bahcall and Oh 1996, Borgani et al. 1997b, Watkins 1997). In this paper
we use the results obtained by Watkins (1997). He developed a
likelihood method for estimating the rms peculiar velocity of clusters
from line-of-sight velocity measurements and their associated errors.
This method was applied to two observed samples of cluster peculiar 
velocities: a sample known as the SCI sample (Giovanelli et al. 1997) and 
a subsample of the Mark III catalog (Willick et al. 1997).  Watkins (1997) 
found the rms one-dimensional cluster peculiar velocity of 
$265^{+106}_{-75}$ km s$^{-1}$, which
corresponds to the three-dimensional rms velocity of 
$459^{+184}_{-130}$ km s$^{-1}$.
 
Figure~4 shows the limits for $\sigma_8 \Omega_0^{0.6}$ in different models, 
assuming that the observed cluster sample studied by Watkins (1997) 
corresponds to the model clusters with 
a mean cluster separation $d_{cl}\sim 30h^{-1}$ Mpc 
($n_{cl}=3.70 \times 10^{-5} h^3$ Mpc$^{-3}$). In the model (1), the rms 
peculiar velocity of clusters with a
separation $d_{cl}=30h^{-1}$ Mpc is $459^{+184}_{-130}$ km s$^{-1}$, when 
$\sigma_8=(0.33^{+0.13}_{-0.09}) f^{-1}(\Omega_0)$. 
For $\Omega_0=0.3$ and $\Omega_0=0.4$, we obtain  
$\sigma_8=0.68^{+0.27}_{-0.19}$ and $\sigma_8=0.57^{+0.22}_{-0.16}$, 
respectively. For the power spectrum (2), we found that 
$\sigma_8=(0.37^{+0.15}_{-0.10} ) f^{-1}(\Omega_0)$. Therefore, for  
$\Omega_0=0.3$ and $\Omega_0=0.4$, $\sigma_8=0.76^{+0.31}_{-0.21}$ and 
$\sigma_8=0.64^{+0.26}_{-0.17}$, respectively.

Now we can compare the observational constraints obtained by studying
the mass function and peculiar velocities of clusters of galaxies. 
In the model (1) for $\Omega_0=0.4$, the mass function and the peculiar 
velocities of clusters are consistent with the observed data, for the 
small window of $\sigma_8=0.71 - 0.78$. For $\Omega_0>0.6$, the 
observed mass function and the peculiar velocities of clusters are not 
consistent with each other. Either the observed mass function of clusters 
is overestimated or the peculiar velocities of clusters are underestimated. 
Therefore, in the model (1), the mass function and the peculiar velocities 
of clusters are consistent with observations only if $\Omega_0<0.6$. In 
the second model, the permitted window in the ($\Omega_0,\sigma_8$) plane 
is larger. For $\Omega_0=0.4$, the mass function and 
the peculiar velocities are consistent with the observed data if
$\sigma_8=0.72-0.88$. For example, for $\Omega=0.4$ and $\sigma_8=0.75$, 
the number density of clusters with mass 
$M_{1.5}=4\times 10^{14} h^{-1} M_{\odot}$ is 
$n(>M)=2.7 \times 10^{-6} h^3$ Mpc$^{-3}$, and the rms peculiar velocity of
clusters with a mean separation $d_{cl}=30h^{-1}$ Mpc is $530$ km s$^{-1}$.

\begin{figure*}
\centering
\begin{picture}(300,350)
\includegraphics{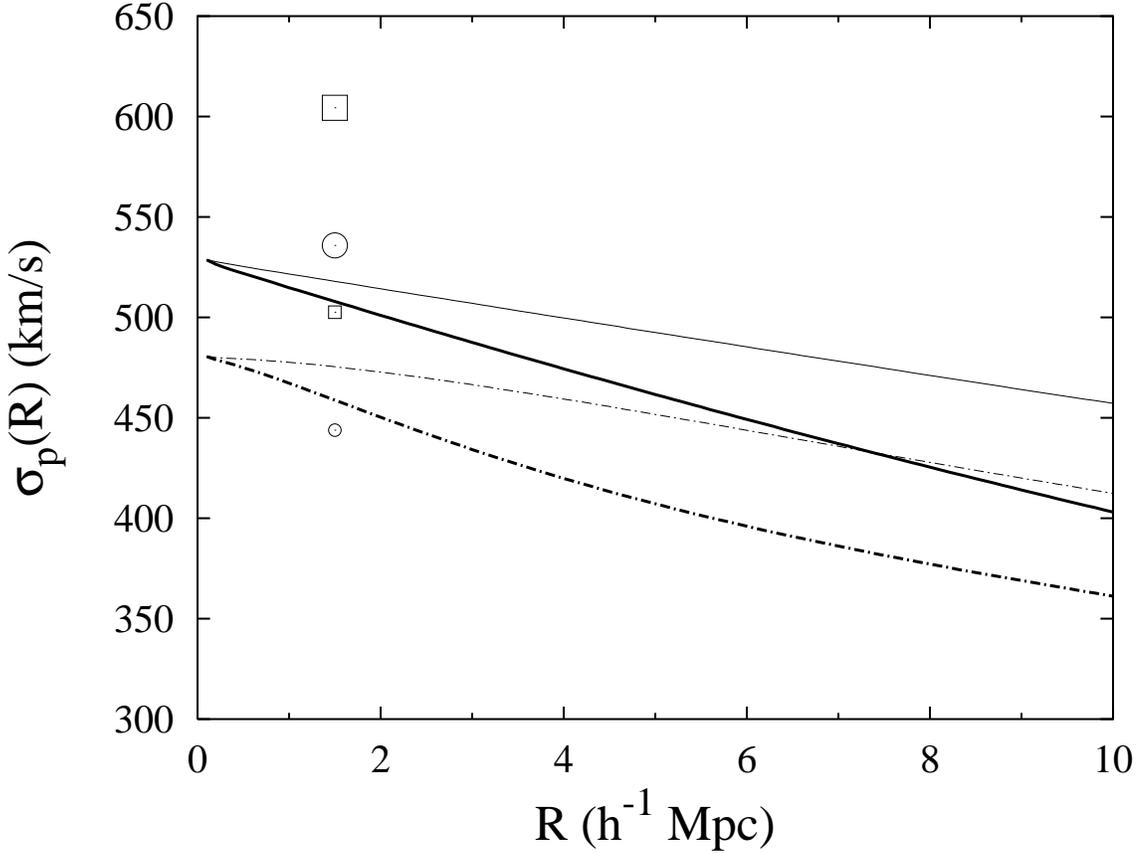}
\end{picture}
\caption {The rms peculiar velocity of peaks, $\sigma_p(R)$, 
for the power spectra (1) (heavy solid line) and (2) (heavy
dot-dashed line) for $\Omega_0=0.3$ and $\sigma_8=0.9$. The light
curves show the corresponding rms peculiar velocity $\sigma_v(R)$ for
the same models. The small and large squares demonstrate the rms peculiar
velocity of clusters in the model (1) with mean separations 
$d_{cl}=10h^{-1}$ Mpc and $d_{cl}=30h^{-1}$ Mpc, respectively. The small 
and large circles show the corresponding rms peculiar velocities in the 
model (2).}
\end{figure*}

We investigated also the linear theory predictions for peculiar 
velocities of peaks in the Gaussian field. 
The linear rms velocity fluctuation on a given scale $R$ can be expressed 
as $$
\sigma_v(R)=H_0 f(\Omega_0) \sigma_{-1} (R),
\eqno(8)
$$
where $\sigma_j$ is defined for any integer $j$ by 
$$
\sigma_j^2={1 \over 2\pi^2} \int P(k) W^2(kR) k^{2j+2} dk.
\eqno(9)
$$
Bardeen et al. (1986) showed that the rms peculiar velocity at peaks of
the smoothed density field differs systematically from $\sigma_v(R)$,
and can be expressed as
$$
\sigma_p(R)=\sigma_v(R) \sqrt{1 - \sigma_0^4/\sigma_1^2 \sigma_{-1}^2} .
\eqno(10)
$$
In this approximation, the rms velocities of the peaks do not depend on
the height of the peaks.

Figure~6 shows the rms peculiar velocities of peaks, $\sigma_p(R)$, for
the power spectra (1) and (2) for $\Omega_0=0.3$ and $\sigma_8=0.9$. 
We have used the top-hat window function. For comparison, we show also the 
rms peculiar velocity $\sigma_v(R)$ for the same models. For the 
cluster radius $R=1.5h^{-1}$ Mpc, $\sigma_p$ is 
lower than $\sigma_v$ about $\sim 2$\% and $\sim 3.5$\% for models (1)
and (2), respectively. On larger scales, the difference between
$\sigma_p$ and $\sigma_v$ increases. For comparison, we show in 
Figure~6 the rms peculiar velocity of clusters with mean separations 
$d_{cl}=10h^{-1}$ Mpc and $d_{cl}=30h^{-1}$ Mpc, derived by using N-body 
simulations. The rms peculiar velocity of clusters with a separation 
$d_{cl}=10h^{-1}$ Mpc is similar to the linear theory expectations at the 
scale $R\sim 1.5h^{-1}$ Mpc. (It is slightly smaller ($\sim 2$\%) than 
$\sigma_p$ at the radius $R=1.5h^{-1}$ Mpc. This small difference is probably
caused by smoothing inherent to particle-mesh method). 

The peculiar velocity of rich clusters is higher than that predicted by the 
linear approximaton (10). In the model (1), the peculiar velocity of 
clusters with a mean separation $d_{cl}=30h^{-1}$ Mpc is $\sim 19.0$\% 
higher than $\sigma_p$ at the radius $R=1.5h^{-1}$ Mpc. In the model (2), 
the peculiar velocity of these clusters is $\sim 16.8$\% 
higher than that predicted by the linear theory. These results are in good 
agreement with the results obtained by Colberg et al. (1998). They compared 
the peculiar velocities of galaxy clusters with $\sigma_p(R)$ at  
larger radii, $R\sim 8h^{-1}$ Mpc, and for this reason found a larger 
difference ($\sim 40$\%) between the rms peculiar velocity of galaxy 
clusters and $\sigma_p$. 

\begin{figure*}
\centering
\begin{picture}(300,550)
\includegraphics{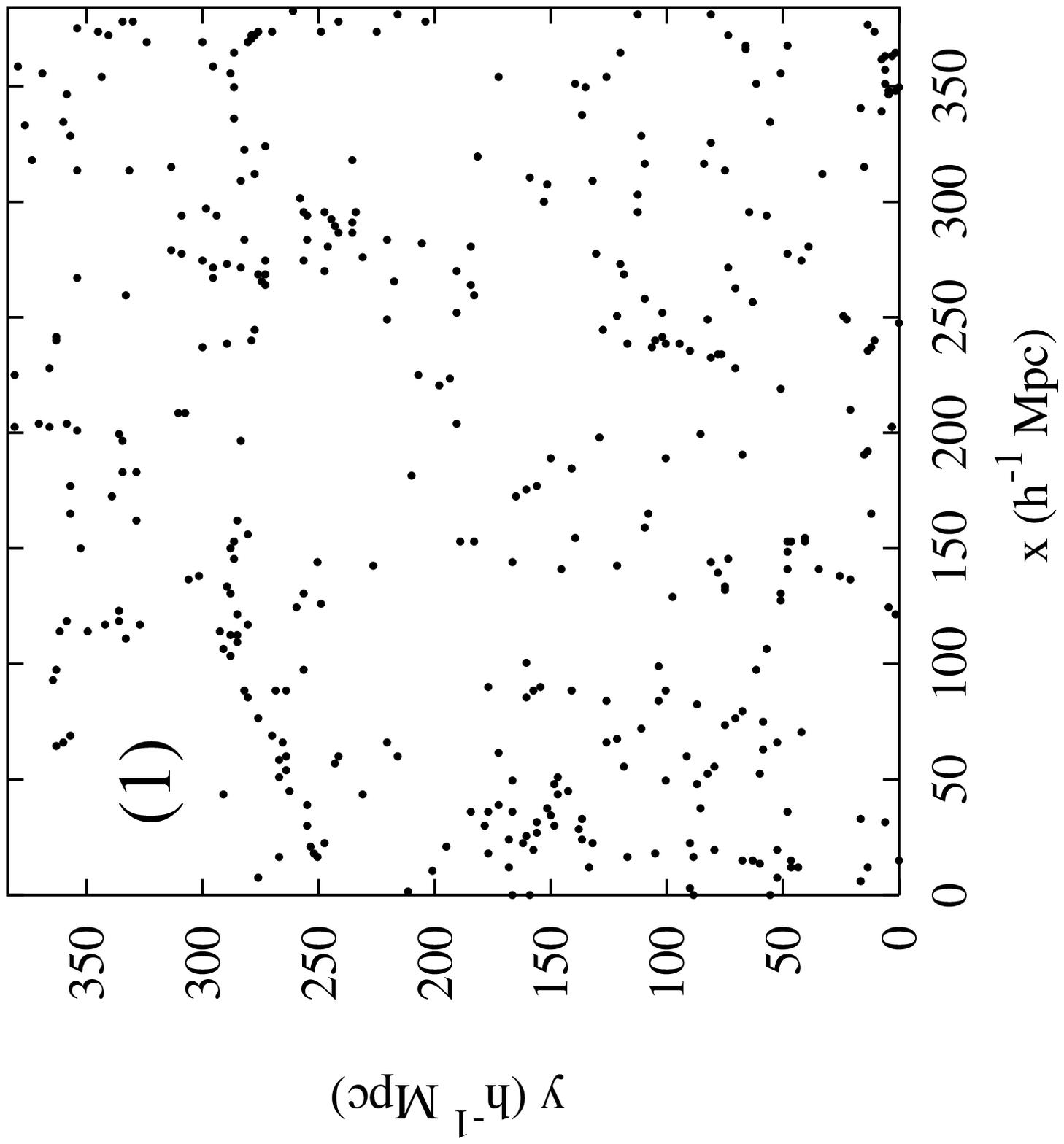}
\includegraphics{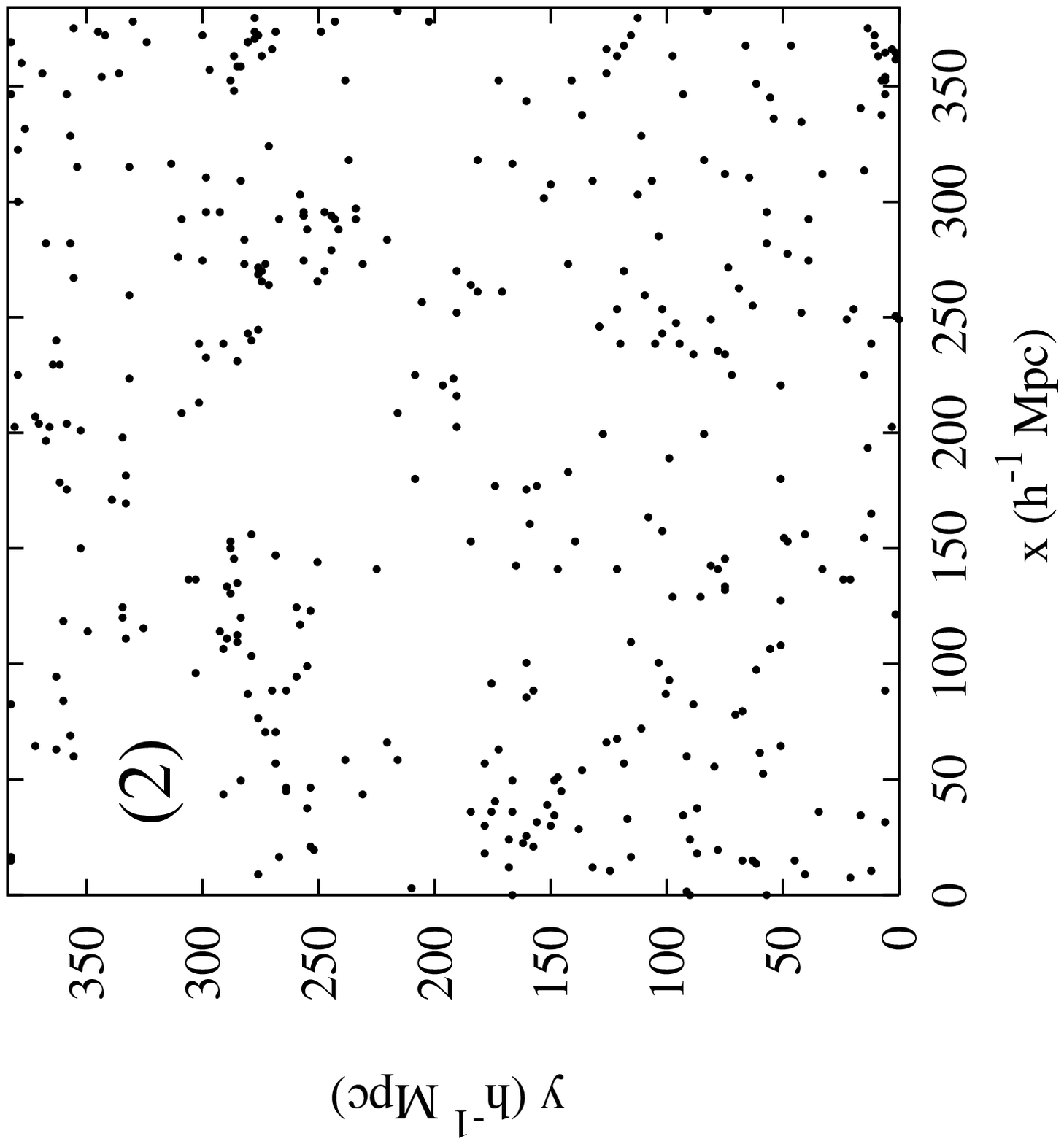}
\end{picture}
\caption {Panel (a) shows the distribution of clusters in a 
$96h^{-1}$ Mpc thick slice in the model (1). Panel (b) shows the
distribution of clusters in the same slice in the model (2). The
distribution is shown for the clusters with a mean separation
$d_{cl}=34h^{-1}$ Mpc.}
\end{figure*}

In Figure~5, we showed that during the evolution between
$\sigma_8=0.5$ and $\sigma_8=0.8$, the rms peculiar velocities 
of clusters with a mean separation $d_{cl}<80h^{-1}$ Mpc evolve as 
expected by the linear theory, $v_{cl} \sim \sigma_8$. The difference 
between the linear approximation (10) and the peculiar velocities of the
$d_{cl}=30h^{-1}$ Mpc clusters must therefore arise when $\sigma_8<0.5$. 
Further study is needed to investigate the nonlinear evolution of peculiar 
velocities of clusters in more detail.

\sec{THE POWER SPECTRUM OF CLUSTERS OF GALAXIES}
\begin{figure*}
\centering
\begin{picture}(300,490)
\includegraphics{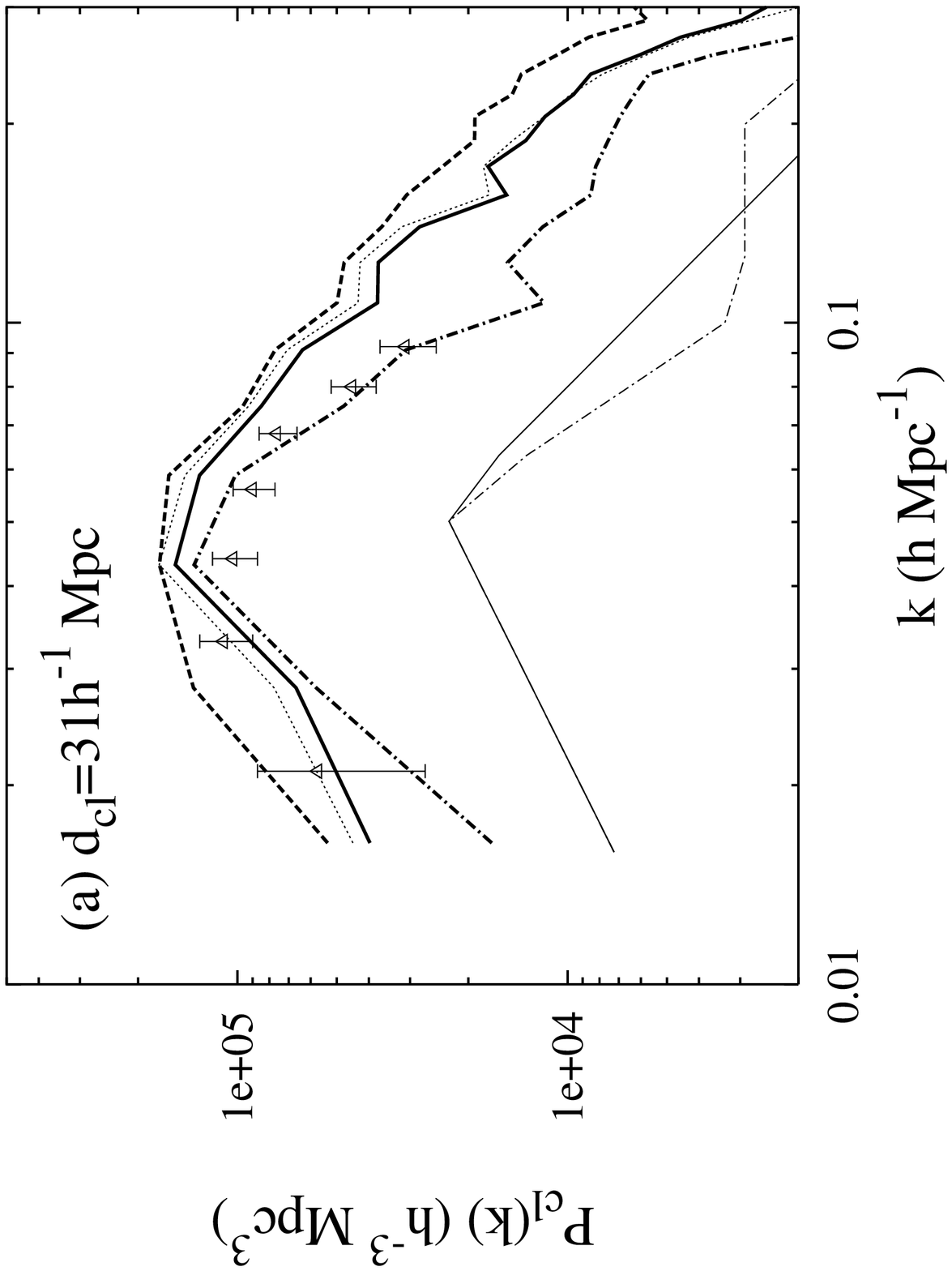}
\includegraphics{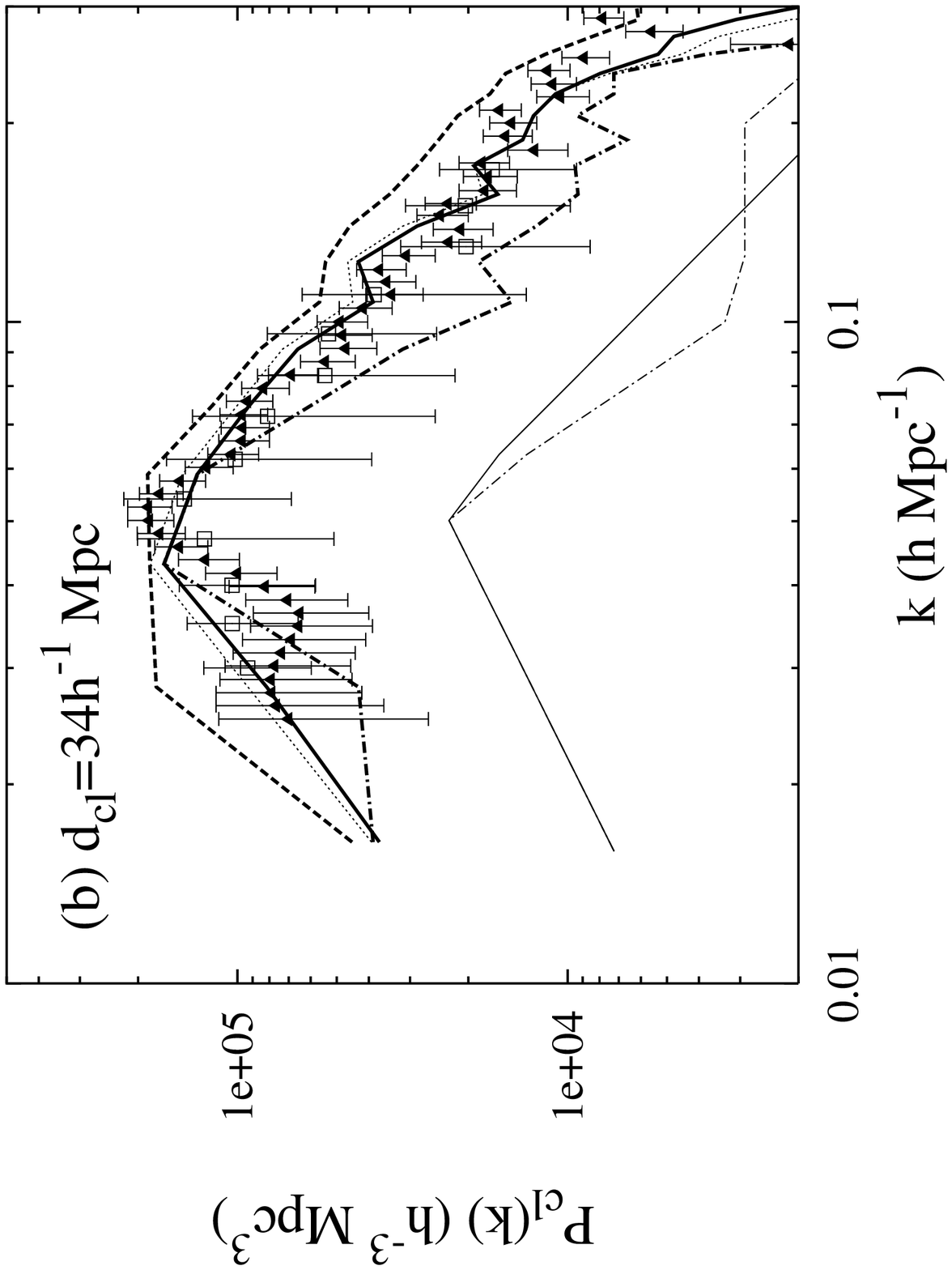}
\end{picture}
\caption {The redshift-space power spectrum of clusters
in the model (1) for $\sigma_8=0.8$ (solid lines) and in the model (2) 
for $\sigma_8=0.84$ (dot-dashed lines). The heavy 
curves show the power spectra of clusters and the light curves the 
corresponding linear power spectra of matter fluctuations. 
To transform the clusters to the redshift space we have assumed that 
$\Omega_0=0.3$. The dotted line demonstrates the redshift-space power 
spectrum of clusters in the model (1) for $\Omega=1$. 
The dashed line represents the power spectrum of clusters in the 
model (1) for $\sigma_8=0.5$. (a) The power spectrum of model clusters
and APM clusters (open triangles) with a mean separation
$d_{cl}=31h^{-1}$ Mpc. (b) The power spectrum of model clusters and
Abell clusters with a separation $d_{cl}=34h^{-1}$ Mpc. Filled triangles 
and open squares show the power spectrum of Abell clusters as determined 
by Einasto et al. (1997a) and Retzlaff at al. (1998), respectively.}
\end{figure*}

Figure~7 shows the spatial distribution of clusters with a mean
separation $d_{cl}=34h^{-1}$ Mpc in our models. Figure~7a shows the 
distribution of clusters for the initial power spectra (1) and Figure~7b 
for the initial spectra with a primordial feature at wavelengths
$\lambda \sim 30-60h^{-1}$ Mpc (equation [2]). The phases of 
the initial density fluctuations were chosen to be same, and therefore, we see 
directly the influence of the initial power spectrum. At a first look, 
the distribution of clusters in different models is rather similar. 
However, we see that the shape of superclusters in different models 
is slightly different. In the model (1), the clusters inside superclusters 
are more concentrated, while in the model (2), superclusters are larger 
and the clusters within superclusters are more disperse (compare, e.g., the 
large superclusters in the upper-left corner). The bump in the initial power 
spectrum at $\lambda \sim 30-60h^{-1}$ Mpc influences the distribution 
of clusters inside superclusters.

Figure~8 demonstrates the redshift-space power spectrum of clusters,
$P_{cl}(k)$. The clusters were determined in the simulation 
at the moment when $\sigma_8=0.8$ and $\sigma_8=0.84$, in the models (1) 
and (2), respectively. To calculate the power spectrum we transformed the
cluster positions to the redshift space, determined the density field on a
$128^3$ grid using the CIC scheme and calculated its Fourier components,  
subtracting the shot noise term.

To transform the clusters from the real space to the redshift space, we 
determined the peculiar velocities of clusters in the simulations with 
$\Omega=1$ and used the linear scaling, $v_{cl} \sim f(\Omega_0)$, for 
$\Omega_0=0.3$. In this case the peculiar velocities of 
clusters are consistent with observations for $\sigma_8 \sim 0.8$ 
(see Figure 5). However, $P_{cl}(k)$ is not very sensitive to this 
assumption. In Figure~8 we show also the power spectrum of clusters in
the redshift space for the model (1) for $\Omega=1$, where the peculiar 
velocities of clusters are severly overestimated, 
$v_{cl} \sim 1100$ km s$^{-1}$. In this model $P_{cl}(k)$ at 
wavenumbers $k<0.1h^{-1}$ Mpc is $\sim 10$\% higher 
than for $\Omega_0=0.3$.

We investigated also the cluster power spectrum in the model (1) for 
$\sigma_8=0.5$ (here we used $\Omega=1$). We found that during the 
evolution between $\sigma_8=0.5$ and $\sigma_8=0.8$, the power spectrum of 
clusters with a mean separation $d_{cl}\leq 30h^{-1}$ Mpc is almost 
unchanged. For richer clusters, $P_{cl}(k)$ somewhat decreases 
($\sim 14$\% and $\sim 29$\% for the clusters with $d_{cl}=31h^{-1}$ Mpc 
and $d_{cl}=34h^{-1}$ Mpc, respectively.) This effect is probably caused 
by merging of very rich clusters. Further study (e.g. numerical 
simulations with higher dynamical range) is needed to determine whether 
this is a real effect for the model (1), or a numerical effect due to the 
limited dynamical range of the N-body simulations. 

Let us compare the power spectra of clusters in our models with the
observed power spectra of the APM and Abell clusters. Figure~8a shows the 
power spectrum of model clusters with a mean separation $d_{cl}=31h^{-1}$ Mpc. 
For comparison, we show the power spectrum of the observed APM clusters 
determined by Tadros, Efstathiou \& Dalton (1998). They analyzed the 
redshift survey of 364 clusters described by Dalton et al. (1994). The 
mean intercluster separation of the APM clusters is $d_{cl}\sim 31 h^{-1}$Mpc 
(Dalton et al. 1994). Figure~8a shows that the power spectrum of the APM 
clusters is in good agreement with the power spectrum of clusters predicted 
in the model (2). In the model (1), the power spectrum of clusters is higher 
than observed (factor of $\sim 1.8$ at $k\sim 0.07-0.08h$ Mpc$^{-1}$). 

Figure~8b demonstrates the power spectrum of model clusters and the Abell 
clusters with a mean separation $d_{cl}=34h^{-1}$ Mpc. We show the power 
spectra of the Abell clusters determined by Einasto et al. (1997a) and 
Retzlaff et al. (1998). Einasto et al. (1997a) determined the power spectrum 
of the Abell clusters from the correlation function of clusters, while 
Retzlaff et al. (1998) estimated the power spectrum directly from the Fourier 
space. We see that the power spectra estimated by different methods are 
consistent with each other. However, the error bars measured by Retzlaff 
et al. (1998) are much larger than calculated by Einasto et al. (1997a) and 
probably underestimated in the latter case (see Retzlaff et al. 1998, 
Einasto et al. 1997a for details). Figure~8b shows that the power spectrum 
predicted in the model (1) for $\sigma_8=0.8$ is consistent with the observed 
power spectrum of the Abell clusters. For $\sigma_8=0.5$, the amplitude of 
density fluctuations at wavenumbers $k>0.1h$ Mpc$^{-1}$ is higher 
than observed. The power spectrum of clusters in the model (2) is 
consistent with the observed power spectrum measured by 
Retzlaff et al. (1998) within the uncertainties. 

We investigated also the relation between the power spectrum of
clusters and the power spectrum of matter fluctuations. 
During the evolution the power spectrum of clusters, $P_{cl}(k)$, is
almost unchanged, while the power spectrum of matter fluctuations
evolves as $P(k) \sim \sigma_8^2$ in the linear regime. In this case, 
the bias parameter $b^2_{cl}(k)=P_{cl}(k)/P(k) \sim \sigma_8^{-2}$ and 
therefore, in order to compare different models at different moments, 
it is reasonable to express the parameter $b_{cl}$ in terms of $\sigma_8$. 
We examined the bias parameter $b_{cl}$ at the wavenumber interval 
$k \sim 0.06-0.08h$ Mpc$^{-1}$. In the model (1) for $\sigma_8=0.8$, we 
found that $b_{cl}=2.30/\sigma_8$ and $b_{cl}=2.40/\sigma_8$ for the 
clusters with $d_{cl}=31h^{-1}$ Mpc and $d_{cl}=34h^{-1}$ Mpc, respectively. 
(For $\sigma_8=0.5$, as the cluster power spectrum is somewhat larger, the 
$b_{cl}=2.45/\sigma_8$ and $b_{cl}=2.70/\sigma_8$, respectively). In the 
model (2) for $\sigma=0.84$, we obtain that $b_{cl}=2.25/\sigma_8$ and 
$b_{cl}=2.50/\sigma_8$, respectively. In this model, the bias parameter 
is similar to the model (1). The relation between the distribution of 
matter density and of clusters of different richness is studied in more 
detail in the paper by Gramann \& Suhhonenko (1998). 

\sec{THE CORRELATION FUNCTION OF CLUSTERS OF GALAXIES}
 
In this Section we examine the spatial two-point correlation function 
of clusters, $\xi_{cl}(r)$, in our models. To determine the correlation 
function, we first tranformed cluster positions to the redshift space and then 
determined the correlation function of clusters by pair counting. 

\begin{figure*}
\centering
\begin{picture}(300,490)
\includegraphics{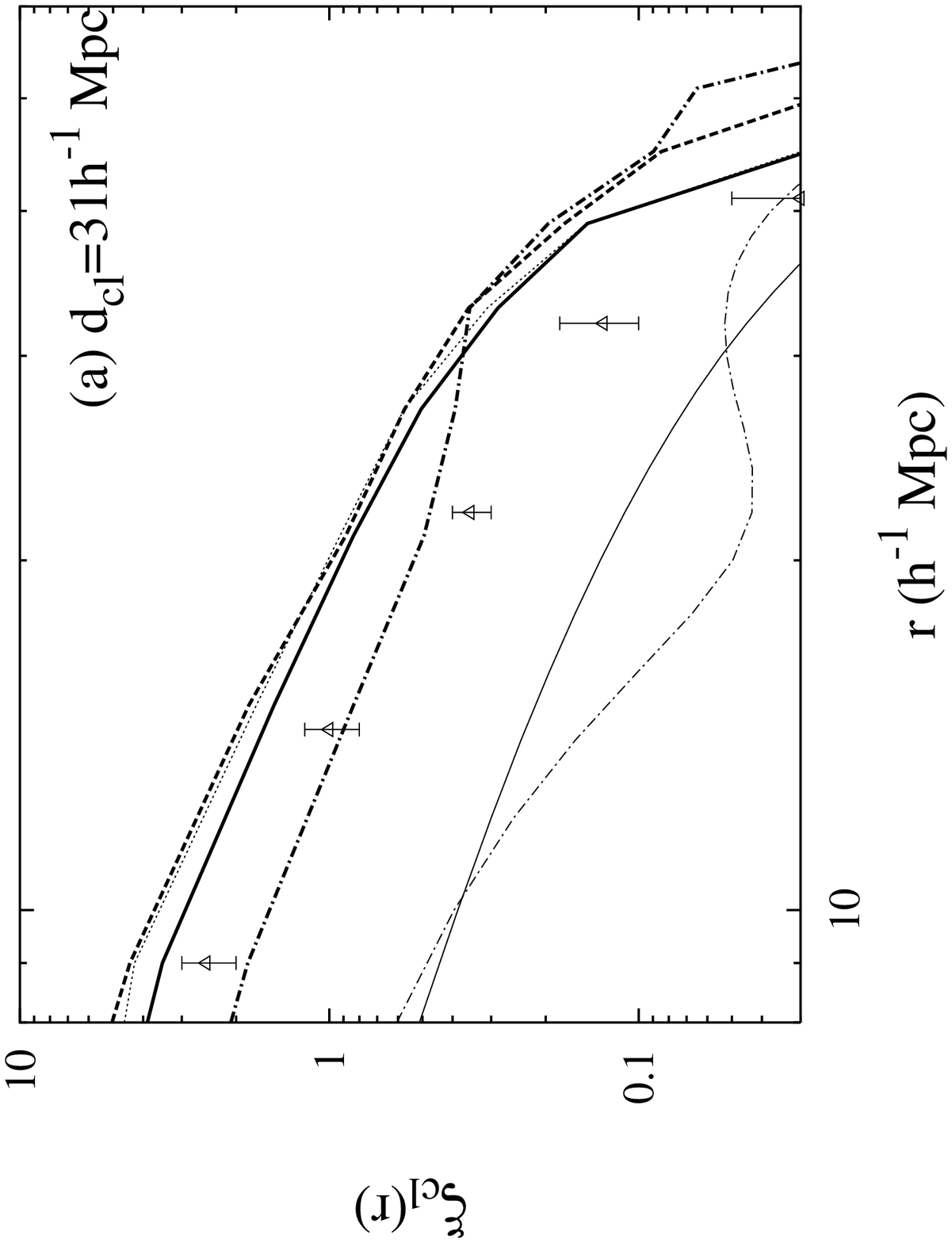}
\includegraphics{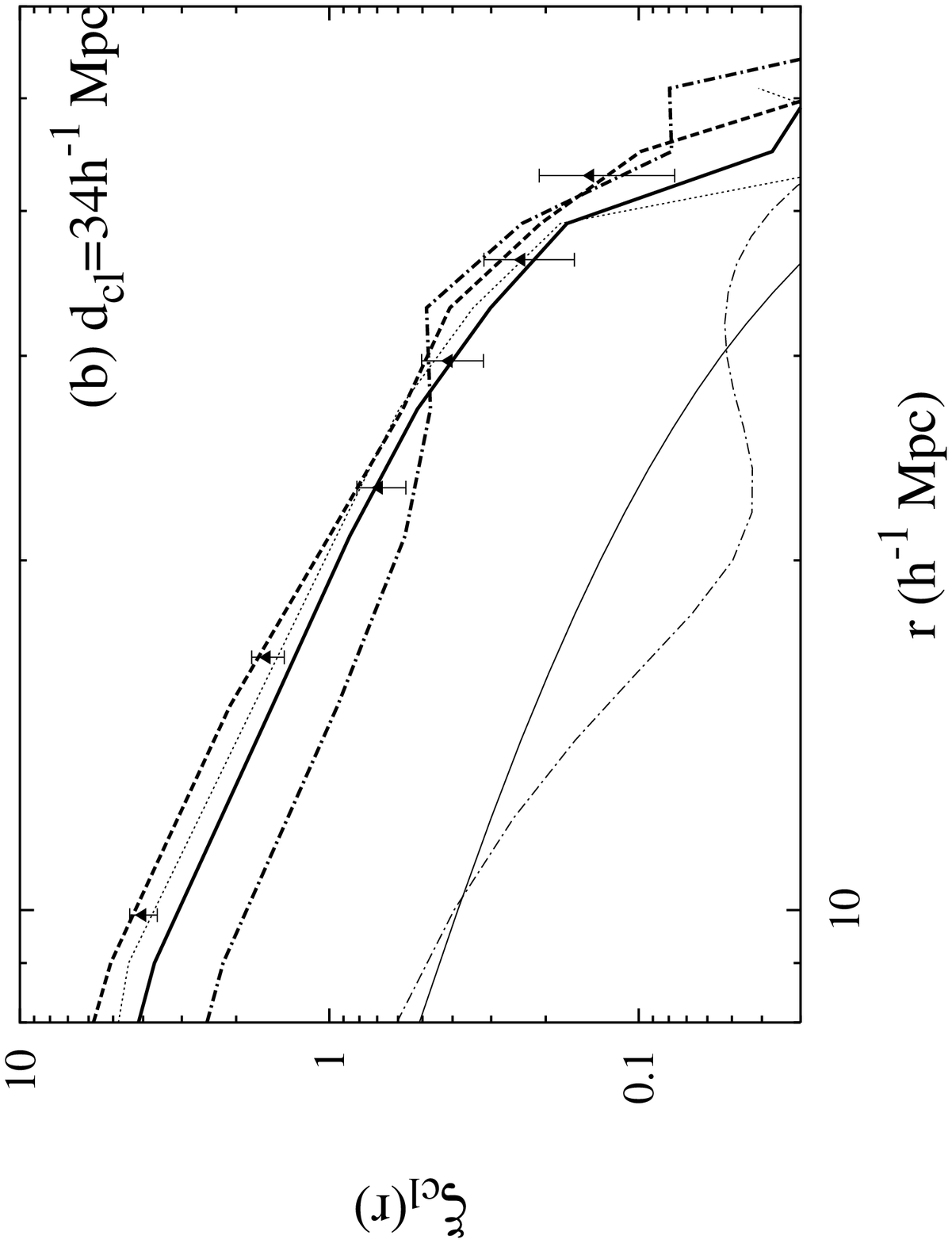}
\end{picture}
\caption {The redshift-space correlation function of clusters
in the model (1) for $\sigma_8=0.8$ (solid lines) and in the model (2) 
for $\sigma_8=0.84$ (dot-dashed lines). The heavy  
curves show the correlation function of clusters and the light 
curves the corresponding linear correlation function of 
matter fluctuations. To transform the clusters to the redshift space we 
have assumed that $\Omega_0=0.3$. The dotted line demonstrates the 
redshift-space correlation function of clusters in the model (1) for 
$\Omega=1$. The dashed line represents the correlation function of 
clusters in the model (1) for $\sigma_8=0.5$.  (a) The correlation function 
of model clusters and APM clusters (open triangles) with a mean separation
$d_{cl}=31h^{-1}$ Mpc. (b) The correlation function of model clusters and
Abell clusters (filled triangles) with a separation $d_{cl}=34h^{-1}$ Mpc.} 
\end{figure*}

Figure~9 shows the redshift-space correlation function of clusters in 
the model (1) for $\sigma_8=0.8$ and in the model (2) for $\sigma_8=0.84$. 
The peculiar velocities were calculated for the density parameter 
$\Omega_0=0.3$. We investigated also the $\Omega=1$ model for 
velocities and found that in this case $\xi_{cl}(r)$ at radii 
$r=10-20h^{-1}$Mpc is $\sim 16$\% higher than for $\Omega_0=0.3$. 
Figure~9 shows also the correlation function in the model (1) for 
$\sigma_8=0.5$. During the evolution in the model (1), the correlation 
function of clusters, as the power spectrum of clusters, somewhat decreases 
($\sim 22$\% and $\sim 38$\%, at $r\sim 15h^{-1}$ Mpc, for the 
$d_{cl}=31h^{-1}$ Mpc and $d_{cl}=34h^{-1}$ Mpc clusters, respectively).
This effect is probably caused by merging of rich clusters.

Now we can compare the correlation function of clusters predicted in
the models with the observed correlation function of the APM and Abell
clusters. Figure~9a demonstrates the correlation function of model
clusters and of the APM clusters with a mean separation $d_{cl}=31h^{-1}$ Mpc.
The correlation function of the APM clusters has been determined by 
Dalton et al. (1994). On small scales the correlation function in the model
(2) is in good agreement with the correlation function of the APM clusters.
The correlation function of the APM clusters is equal to unity at a pair
separation $r_0=14.3\pm 1.75h^{-1}$ Mpc (Dalton et al. 1994). In the
model (2) we find that $r_0=14 \pm 1h^{-1}$ Mpc. 

On larger scales, $\xi_{cl}(r)$ in the model (2) is larger than that
measured by Dalton et al. (1994) for the APM clusters. On the other hand, 
the power spectrum of clusters on large spatial scales in the model (2) 
is in good agreement with the power spectrum of the APM clusters determined 
by Tadros, Efstathiou \& Dalton (1998) (Figure~8a). This comparision 
suggests that the correlation function of clusters determined by 
Dalton et al. (1994) may be underestimated at large separations 
$r>20h^{-1}$ Mpc. In the model (1), $\xi_{cl}(r)$ is higher than 
observed on all scales. In this model $r_0=19 \pm 1h^{-1}$ Mpc.

Figure~9b shows the spatial correlation function of clusters with a 
separation $d_{cl}=34h^{-1}$ Mpc. For comparison, we present the correlation
function of the Abell clusters determined by Einasto et al. (1997c). On
small scales the correlation function of the Abell clusters is higher than
that observed for the APM clusters. We find that $r_0=20\pm 3h^{-1}$ Mpc for
the Abell clusters. This effect is partly due to differences in the
number densities of the APM and Abell clusters. However, our models predict 
that the correlation function of clusters with a mean separation 
$d_{cl}=31h^{-1}$ Mpc and $d_{cl}=34h^{-1}$ Mpc is very similar. 
(In the models (1) and (2), the correlation length of clusters with 
$d_{cl}=34h^{-1}$ Mpc is $r_{0}=20 \pm 1h^{-1}$ Mpc and 
$r_0=15\pm 1h^{-1}$ Mpc, respectively.) The correlation function of the
Abell clusters at separations $r<25h^{-1}$ Mpc is probably overestimated 
due to the projection and selection biases known to affect clustering 
in the Abell cluster catalogues (e.g. Sutherland 1988). On larger 
separations these effects are not so important. The correlation function of 
clusters in the models (1) and (2) is consistent with the observed
correlation function of the Abell clusters on separations $r>25h^{-1}$ Mpc.
We examined also the radius at which the cluster correlation function
$\xi_{cl}(r_1)=0$. We found that $r_1=54\pm3h^{-1}$ Mpc and 
$r_1=60\pm3h^{-1}$ Mpc, in the models (1) and (2), respectively. For the 
Abell clusters, the parameter $r_1=50\pm10h^{-1}$ Mpc.

Thus, the model (2) fits the correlation function of the APM clusters on 
small scales and the correlation function of Abell clusters on large scales. 
This model predicts that there is a bump in the correlation function of 
clusters at separations $r\sim 20-35h^{-1}$Mpc. Available data are
insufficient to confirm or to rule out this interesting possibility. 
Accurate measurements of the correlation function of clusters at these 
distances can serve as a discriminating test for this model. 

\sec{CONCLUSIONS}

In this paper, we have examined the properties of clusters of
galaxies in two cosmological models. In the first model, the initial power
spectrum was chosen in the form $P(k) \propto k^{-2}$ at the wavelengths
$\lambda<120h^{-1}$ Mpc (equation 1). In the second model, we
assumed that the initial power spectrum contains a primordial feature 
at the wavelengths $\lambda \sim 30-60h^{-1}$ Mpc (equation 2).
The density fluctuations at these wavelengths influence the distribution 
of clusters inside superclusters. In the model (2), superclusters 
are larger and the clusters inside superclusters are not so concentrated than 
in the model (1) (see Figure 7). We investigated the mass 
function, peculiar velocities, the power spectrum and the correlation 
function of clusters in both models for different values of $\Omega_0$ and
$\sigma_8$. Below, we briefly summarize the results obtained.

(1) The mass function of clusters of galaxies in models (1) and (2), 
when compared at a same values of $\Omega_0$ and $\sigma_8$, is very 
similar for smaller masses $M \leq 4 \times 10^{14} h^{-1} M_{\odot}$. 
For larger masses the mass function in the model (2) is steeper than in 
the model (1). For $\Omega_0=0.3$ and $\Omega_0=0.4$, the mass function of
clusters in both our models is consistent with observations, if 
$\sigma_8=0.90\pm 0.12$ and $\sigma_8=0.80 \pm 0.09$, respectively. 

(2) The rms peculiar velocities of clusters in the model (2) are $\sim 12$\% 
smaller than in the model (1), when compared at the same values of 
$\Omega_0$ and $\sigma_8$. In the model (1), the rms peculiar velocity of 
clusters is consistent with observations if 
$\sigma_8=(0.33^{+0.13}_{-0.09}) \Omega_0^{-0.6}$. In this model, the 
mass function and the peculiar velocities of clusters are consistent with 
observations only if $\Omega_0<0.6$. For $\Omega_0=0.4$, the mass function 
and the peculiar velocities are consistent with the observed data if 
$\sigma_8=0.71 - 0.78$. 
In the model (2), the rms peculiar velocity of clusters is consistent
with observations if $\sigma_8=(0.37^{+0.15}_{-0.10}) \Omega_0^{-0.6}$ and
the permitted region in the ($\Omega_0,\sigma_8$) plane is larger. 
For $\Omega_0=0.4$, the mass function and the peculiar 
velocities are consistent with the observed data if $\sigma_8=0.72-0.88$.

(3) The redshift-space power spectrum of clusters in the model (2)
is in good agreement with the observed power spectrum of the APM clusters.
The power spectrum of clusters in this model is also consistent with the 
observed power spectrum of the Abell clusters within uncertainties.
In the model (1), the power spectrum of clusters is higher than
observed for the APM clusters (factor of $\sim 1.8$ at $k\sim
0.07-0.08h$ Mpc$^{-1}$).

(4) The redshift-space correlation function of clusters in the model (2) 
is consistent with the correlation function of the APM clusters at small 
distances $r<25h^{-1}$ Mpc. At larger separations the cluster correlation 
function in this model is consistent with the correlation function as
derived for the Abell clusters. In the model (1), the correlation 
function of clusters on small distances is higher than observed for the
APM clusters.

Therefore, in many aspects the power spectrum of density fluctuations 
in the model (2) fits the observed data better than the simple power 
law model (1). The superclusters of galaxies in the Universe are probably 
more disperse as predicted in the model (2) and not so concentrated as 
predicted in the model (1). Observed data are not sufficient to examine
the power spectrum of density fluctuations at wavelengths 
$\lambda \sim 30-120h^{-1}$ Mpc in more detail, but our study suggests
that probably at these wavelengths the initial power spectrum is not
a featureless simple power law. 

We examined also the linear theory predictions for peculiar 
velocities of peaks in a Gaussian field and compared these with the
peculiar velocities of clusters in N-body simulations. We determined
the clusters as the maxima of the density field smoothed on the scale
$R\sim 1.5h^{-1}$ Mpc and defined their peculiar velocities using the
same smoothing scale as for the density field. In this way, the sizes
of all clusters are the same and do not depend on the richness of the
cluster. Our study shows that in this case the rms peculiar velocities 
of clusters increase with cluster richness. The rms peculiar velocity 
of small clusters is similar to the linear theory expectations, while 
the rms peculiar velocity of rich clusters is higher than that predicted 
in the linear theory.

\sec*{ACKNOWLEDGEMENTS}

We thank E. Saar, J. Einasto and J. Colberg for useful discussions. This 
work has been supported by the ESF grant 97-2645.

\vfill

\end{document}